\documentclass[inbook,preprint]{egs} 

\usepackage{wku}

\begin{document}

\makeatletter
\@mathmargin1pt
\makeatother

\title{On the averaged quantum dynamics by\\ white-noise Hamiltonians
       with and without dissipation}

\author[$\,$1,*]{Werner Fischer}
\author[$\,$1]{Hajo Leschke} 
\affil[1]{Institut f\"ur Theoretische Physik,
          Universit\"at Erlangen-N\"urnberg,  
          Staudtstra{\ss}e~7, 91058 Erlangen, Germany}
\author[$\,$1,2]{Peter M\"uller}
\affil[2]{Institut f\"ur Theoretische Physik,
          Georg-August-Universit\"at, 37073 G\"ottingen, Germany}

\booktitle{Dedicated to Wolfgang Kundt on the occasion of his 
           65$^{\,\mathit{th}}$ birthday}

\firstauthor{Fischer}

\maketitle

\vspace{17pt}

\begin{abstract}
  Exact results are derived on the averaged dynamics of a class of
  random quantum-dynamical systems in continuous space.  Each member
  of the class is characterized by a Hamiltonian which is the sum of
  two parts. While one part is deterministic, time-independent and
  quadratic, the Weyl-Wigner symbol of the other part is a homogeneous
  Gaussian random field which is delta correlated in time, but
  smoothly correlated in position and momentum.  The averaged dynamics
  of the resulting white-noise system
  is shown to be a monotone mixing increasing quantum-dynamical
  semigroup. Its generator is computed explicitly. Typically, in the course of
  time the mean energy of such a system grows linearly to infinity. In
  the second part of the paper an extended model is studied, which, in
  addition, accounts for dissipation by coupling the white-noise
  system linearly to a quantum-mechanical harmonic heat bath. It is
  demonstrated that, under suitable assumptions on the spectral density 
  of the heat bath, the mean energy then saturates for long times.
\end{abstract}


\renewcommand{\thefootnote}{\fnsymbol{footnote}}
\footnotetext[1]{New address: Siemens AG, Semiconductors, Balanstra\ss{}e 73,
  81541 M\"unchen, Germany}


\section{Introduction and summary of results} \label{intro}

Ever since \citet{And58} one-particle quantum Hamiltonians with 
time-inde\-pend\-ent random potentials have been intensively studied as 
models for disordered electronic systems
\citep{BoEnEsKe+84,ShEf84,LiGrPa88,CaLa90,PaFi92}.
In contrast, models with time-dependent random potentials have
attracted less attention. They have been used, for example, to
explain the diffusive behaviour of Frenkel excitons in molecular
crystals \citep{HaRe72}. Subsequently \citet{OvEr74}, \citet{MaPo77}
and \citet{GiMa79} studied lattice models with time-dependent diagonal
and off-diagonal random parts. Here the randomness enters via
time-dependent fluctuating coupling coefficients and serves to mimic
the influence of collective excitations of the crystal.  At high temperatures
the coupling coefficients should be uncorrelated in time. More
precisely, it is justified \citep{OvEr74} in this case to view them as
realizations of a Gaussian white noise. One is then able to derive
exact analytical results on the averaged dynamics of the model.

The simplifying feature of a vanishing correlation time was also
exploited by \citet{JaKu82}. They were concerned with the averaged
dynamics of a point particle moving in continuous space under the
influence of a time-dependent Gaussian random potential which is
uncorrelated in time, but smoothly correlated in position. Their main
findings are a cubic long-time growth for the spatial spreading
of states under the averaged dynamics and a linear increase in time of
the mean energy. This behaviour is strikingly different from the
diffusive spreading of states known from the aforementioned lattice
models, which goes along with a saturating mean energy and may be
interpreted in terms of \emph{Umklapp}-processes. An extension of
\citet{JaKu82} to the case of an additionally present constant
electric field is given in \citet{Jau87}.

A detailed study of models with uncorrelated
time-dependent randomness is not only of interest for understanding
specific problems in condensed-matter physics. 
Indeed, they are of relevance to many different branches of physics.
Accordingly, the appearance of an additive white noise is not necessarily due
to thermal fluctuations. Such models occur, for instance, in the debate on
fundamental issues of quantum mechanics. In this context their
averaged dynamics allows for a 
quantum-mechanical description of classical objects while still
forbidding superpositions of macroscopically distinct state vectors, see
for example \citet{JoZe85}, \citet{GhRiWe86}, \citet{Joo87},
\citet{GhRiWe87}, \citet[p.~201]{Bel87}, \citet{GhRi90},
\citet{Zur91}, \citet{PeSq94}.  In the realm of cosmology they are
sought \citep{Haw82,BaSuPe84,Haw84} to solve the information problem
\citep{Haw76} which arises by the evaporation of black holes.
Furthermore they are met in classical kinetic theories \citep{Spo80}
and in quasi-classical Markovian transport theories
\citep{MaRiSc90,FeGr95,HaJa96}.

Hence it is desirable to provide a systematic discussion of the
stochastic dynamics of a rather general model with uncorrelated
time-dependent randomness. This was initiated in our earlier paper
\citep{FiLeMu94}, where, as a first step, the averaged dynamics of
such a model was studied. Since proofs of the results were only sketched 
there, it is one goal of the present paper to provide detailed proofs.
The class of quantum-dynamical random systems which was introduced by
\citet{FiLeMu94}---see Model \ref{weissrausch} in the next section---is
characterized by a model Hamiltonian in continuous space which is the sum of a
deterministic part and of a random part.  Apart from allowing the
deterministic part to be quadratic in both momentum \emph{and}
position, which may
serve to describe the influence of a constant magnetic field, say,
it is the other main feature of \citet{FiLeMu94} that the random part
may have fluctuations in both position \emph{and} momentum. The fluctuations in
momentum may be viewed as a continuum analogue of the off-diagonal
random parts in the lattice models mentioned above or as a caricature
for inelastic collisions of the particle. From the point of view of
canonical mechanics they are natural anyway.

As to the mathematical description of random operators with
fluctuations in position and momentum, one could think of applying a
factor-ordering prescription to a random field on classical phase
space. Therefore, and in order to clearly isolate quantum effects from
classical ones even before taking expectation values, 
it is only natural \emph{not} to use the conventional 
Hilbert-space formulation of quantum mechanics, but to represent the
quantum system from the outset in phase space. In so doing, we choose
the linear phase-space representation of Weyl, Wigner and Moyal, which
amounts to a totally symmetric operator ordering when going back to
Hilbert space. 

As was already mentioned it is one goal of the present paper to
provide detailed proofs of the results given in \cite{FiLeMu94}.
This is done in Section~\ref{avdyn}. In Statement \ref{tthg} we establish the
semigroup property of the averaged dynamics of Model
\ref{weissrausch}. Its irreversible character is described in
Statement \ref{irreversibel}.  Moreover, the averaged dynamics is
positivity preserving and hence even a Markov semigroup, see Remark
\ref{classinter}.  Statement \ref{ttzer} is concerned with some
geometrical properties of averaged time-evolved states and Statement
\ref{tth} shows that in typical cases the mean energy of Model
\ref{weissrausch} grows linearly to infinity in the course of time.
This unlimited increase of energy occurs
because there is no mechanism built into Model \ref{weissrausch} which
causes energy to dissipate.  It is therefore the second goal of the
present paper to study also an extended version of Model
\ref{weissrausch} in which dissipation is incorporated, too. This is
done in Section~\ref{diss}, where the white-noise system is linearly coupled to
a quantum-mechanical heat bath of independent harmonic oscillators, see Model
\ref{raureidef}. The averaged dynamics of the total system, that is, 
white-noise system plus heat bath, is analyzed
in Subsection \ref{avtot} with the methods developed in
Section~\ref{avdyn}. In Subsection \ref{lang} we eliminate the degrees
of freedom of the heat bath assuming a thermal equilibrium state of
the heat bath. Statement \ref{klimax} at
the end of Subsection \ref{longtime}, which is the main result of the second
part of the paper, shows that the thus obtained reduced averaged
dynamics of Model \ref{raureidef} acquires the desired dissipative
features in the macroscopic limit of the heat bath provided its spectral 
density behaves suitably. The interplay of
noise and dissipation then prevents the mean energy of the white-noise system
from diverging in the long-time limit. The proof of this result
requires some technical preparations which are dealt with in Subsections
\ref{mac} and \ref{longtime}. Section~\ref{inandout} compiles some
references concerning two related 
challenging problems which we do not address otherwise in this paper:
fluctuations around the averaged dynamics and effects of a
non-vanishing correlation time, that is, coloured-noise perturbations.
Finally, for the convenience of the reader, the basic properties of
the Weyl-Wigner-Moyal representation of quantum mechanics are presented
in the Appendix. The reader who is interested in more details is referred
to the (hopefully) up-to-date list of references.


\section{The averaged dynamics by white-noise Hamiltonians} \label{avdyn}

In what follows, we are interested in the non-relativistic quantum
dynamics of a single spinless point  
particle whose configuration space we choose to be, for simplicity, the 
Euclidean line $ \rz\, $. Following Weyl, Wigner and Moyal we will
represent this quantum system in the
two-dimensional symplectic phase space $ \rxr\, $. The generalization
to several Cartesian degrees of freedom is straightforward and will be
dealt with on demand in Section~\ref{diss}.

\subsection{White-noise Hamiltonians} \label{wnham}

In this subsection we define the basic model of this paper.

\begin{model} \label{weissrausch}
  Being the Weyl-Wigner symbol of a Hamiltonian operator, the
  Hamiltonian function
  \begin{equation} \label{ham}
    H + N_{t}\,,
  \end{equation}
  also referred to as white-noise Hamiltonian in the sequel, defines a
  random quan\-tum-dynamical system on phase space $ \rxr\, $.  The
  deterministic part $ H $ of \eqref{ham} is time independent and at
  most quadratic in momentum $ p $ and position $ q $.  The
  time-dependent random part $ N_{t} $ is a homogeneous Gaussian
  random field in $ p $, $ q $ and time $ t $ with zero mean,
  \begin{equation} 
    \overline{N_{t}(p,q)} = 0\,,                        
  \end{equation}
  and covariance function
  \begin{equation} \label{kov}
    \overline{N_{t}(p,q) \,N_{t'}(p',q')} = \delta (t-t')\,
    C(p-p',q-q')\,.
  \end{equation}
  Here we denote the stochastic average over randomness by an overbar,
  $ \delta $ stands for the Dirac delta function and the phase-space
  part $ C $ of the covariance function is assumed to be arbitrarily often
  differentiable.
\end{model}

\noindent
The requirement for $ H $ to be at most quadratic in position and
momentum is necessary for  deriving explicit results on the
averaged dynamics of Model \ref{weissrausch}. On the other hand the
time independence of $ H $ and the homogeneity of $ N_{t} $ are only
assumed for simplicity. If one were interested in treating
multiplicative noise \citep{Hab94}, this could be managed, for
example, by giving up the homogeneity assumption.

We would like to stress that in order for $ C $ to be a covariance
function it needs to be the Fourier transform of a positive measure
on $ \rxr\, $. This point is disregarded by \citet{Jay93}. Therefore
it comes without surprise that some of his results are physically
insignificant---especially those which are in contradiction to ours.

The dynamics generated by the white-noise Hamiltonian (\ref{ham})
is formally described by the stochastic quantum Liouville equation
\citep{Kub63}
\begin{equation}\label{liou}
\partial_{t} w_{t} = - [H, w_{t}] - [N_{t}, w_{t}]
\end{equation}
with initial condition $ w_{0} = w $. 
Here $ w $ is the Wigner density representing the initial state and 
$ [\winzbullet , \winzbullet ] $ denotes the Moyal bracket (\ref{moyal}) 
of two phase-space functions.
Note that the first Moyal bracket $ [H, w_{t}] $ in (\ref{liou})
is in fact a Poisson bracket due to the quadratic nature of $ H $,
cf.\ Remark \ref{moyalbem}.
Being a stochastic partial differential equation, (\ref{liou}) has to be 
supplemented by a prescription of how to interpret the second Moyal 
bracket on its right-hand side, which involves the distribution-valued
white noise $ N_{t} $. 
We  choose a prescription which is quite natural on physical grounds:
The time change of a realistic physical system is 
governed by driving forces with a non-zero correlation 
time. Theoretical models based on stochastic processes which are 
uncorrelated in time may only be used successfully if the correlation 
time of the actual driving forces is much smaller than any other time scale 
inherent to the system. Therefore we interpret (\ref{liou}) in the
Stratonovich sense which---as is supported by the Wong-Zakai-like
theorems in \citet[p.\ 101]{HoLe84}, 
\citet[Chap.\ 5.2.D]{KaSh88} and \citet{BrFl95}---corresponds to considering 
the white-noise limit $ \tau \downto 0 $ of the solution 
$ w_{t}^{(\tau )} $ of the regularized initial-value problem
\begin{equation}
\begin{split} \label{regliou}
&\partial_{t} w_{t}^{(\tau )} = - [H, w_{t}^{(\tau )}] -
[N_{t}^{(\tau )}, w_{t}^{(\tau )}]\,,\\
&w_{0}^{(\tau )} = w\,.
\end{split}
\end{equation} 
In (\ref{regliou}) we have replaced the Gaussian white noise $ N_{t} $ by the 
Gaussian coloured noise $ N_{t}^{(\tau )} $ which vanishes in mean,
$ \overline{N_{t}^{(\tau )}(p,q)} =0 $, and is correlated according to
\begin{equation} \label{regkov}
\overline{N_{t}^{(\tau )}(p,q) \,N_{t'}^{(\tau )}(p',q')}  =
g^{(\tau )}(t-t')\, C(p-p',q-q')\,,
\end{equation}
where $ g^{(\tau )} $ is a correlation function that approaches 
the Dirac delta function in the limit of vanishing correlation time
$ \tau \downto 0 $. For times $ t\ge 0 $ the solution of (\ref{regliou})
can be conveniently expressed  in terms of a time-ordered exponential
\begin{align}
w_{t}^{(\tau )} &= w + \sum_{n=1}^\infty  (-1)^n \int_{0}^t\!\d s_{n}
\int_{0}^{s_{n}}\!\d s_{n-1}\;\dots\int_{0}^{s_{2}}\!\d s_{1}\; 
\notag\\
&\hspace*{3.3cm} \times\bigl[ H + N_{s_{n}}^{(\tau )}, 
[H + N_{s_{n-1}}^{(\tau )},
\ldots  ,[H + N_{s_{1}}^{(\tau )},w]\ldots ]\bigr] \notag\\
& =: \Texp\biggl\{-\int_{0}^{t} \!\d s\;
[ H + N_{s}^{(\tau )}, \winzbullet ]\biggr\} w\,. \label{formal}
\end{align}
We remark that the corresponding solution for times $ t \le 0 $ requires 
the time ordering to be reversed.

\subsection{The averaged dynamics} \label{aver}

The averaged dynamics generated by a white-noise Hamiltonian is the
mapping $\tt$ defined by 
\begin{equation} 
w \mapsto \tt w := \overline{w_{t}} \,,
\end{equation}
where $ \overline{w_{t}}(p,q) := \overline{w_{t}(p,q)} $ denotes the
stochastic mean of the solution of (\ref{liou}) in the Stratonovich 
interpretation. According to the preceding subsection, $\tt$ can be
expressed as the limit
\begin{equation}   \label{ttdef}
\tt = \lim_{\tau \downto 0}\;\overline{\Texp\biggl\{-\int_{0}^{t} \!\d s\;
[ H + N_{s}^{(\tau )}, \winzbullet ] \biggr\}}\,, \qquad   t\ge 0\,.
\end{equation}
The salient feature of the white-noise limit is that it allows to 
describe the averaged dynamics explicitly.

\begin{satz}\label{tthg}
The averaged dynamics of $ H + N_{t} $ is a semigroup of linear 
operators 
\begin{equation} 
\mathcal{T} _{t}= \exp\{ -t(\mathcal{L} + \mathcal{N})\}\,.
\end{equation}
Its generator consists of the Liouvillian
\begin{equation}  \label{genel}
\mathcal{L} := [H,\winzbullet ] =
(\partial_{p}H)\partial_{q} - (\partial_{q}H)\partial_{p} 
\end{equation}
associated with the deterministic part $ H $ of the Hamiltonian
and of the perturbation
\begin{equation}  \label{genen}
\mathcal{N} := \hbar^{-2}\,\bigl\{ C(0,0) - C(-\i\hbar\partial_{q},
\i\hbar\partial_{p})\bigr\}
\end{equation}
which depends only on the stochastic properties of the random part 
$ N_{t} $.
\end{satz} 

\begin{proof}
To perform the stochastic average in (\ref{ttdef}) it is advantageous 
\citep[{}\S $\,$7 (5)]{Kub62} to isolate the unperturbed time evolution
generated by $ \mathcal{L} $ in passing to the Dirac-Dyson 
representation of the time-evolution operator
\begin{equation}\label{wwbild}
\Texp \biggl\{ - \int_{0}^t\!\d s\; \bigl( \mathcal{L} + 
[ N_{s}^{(\tau )} ,\winzbullet ]\,\bigr)\biggr\}
= \emtl \; \Texp \biggl\{ - \int_{0}^t\!\d s\; \esl 
[ N_{s}^{(\tau )} ,\winzbullet ]\, \emsl \biggr\}\,.
\raisetag{-.6ex}
\end{equation}
Next we introduce the family of Gaussian-distributed  random complex 
measures $ \tilde{N}_{t}^{(\tau )}(\d p\d q) $ on phase space arising 
in the spectral representation \citep[Chap.~11, \S~4]{Doo64}
\begin{equation} \label{nspek}
N_{t}^{(\tau )}(p,q) =: \intpr \tilde{N}_{t}^{(\tau )}(\d p'\d q')\;
\e^{\i (pq'-qp')/\hbar}
\end{equation}
of $N_{t}^{(\tau )}$. They have zero mean, 
$ \overline{\tilde{N}_{t}^{(\tau )}(\d p\d q)} = 0 $,
and the covariance measure
\begin{equation} \label{kovmass}
\overline{\tilde{N}_{t}^{(\tau )}(\d p\d q)\,
\tilde{N}_{t'}^{(\tau )}(\d p'\d q')} 
= g^{(\tau )}(t-t')\,\tilde{C}(\d p\d q)\;
\delta_{-p,-q}(\d p'\d q')\,.
\end{equation}
Here $ \delta _{-p,-q}(\d p'\d q') = \delta (p'+p)\delta(q'+q)\d p'\d q'$ 
denotes the Dirac measure with support $ (-p,-q) \in \rxr\, $. 
The covariance function $ C $ appears 
as the symplectic Fourier transform 
\begin{equation} \label{kovkovmass}
C(p,q) = \intpr \tilde{C}(\d p'\d q')\; \e^{\i (pq'-qp')/\hbar}
\end{equation}
of the measure $ \tilde{C} $ on $ \rxr\, $. Inserting 
(\ref{nspek}) into the definition (\ref{moyal}) of the Moyal bracket
one obtains
\begin{equation} 
[ N_{s}^{(\tau )} ,\winzbullet ] = 
\i \intpr \tilde{N}_{s}^{(\tau )}(\d p'\d q')\; \mathcal{F}(p',q')\,.
\end{equation}
Here we have introduced the operator-valued function
\begin{equation} \label{fop}
\rxr \ni (p',q') \mapsto \mathcal{F}(p',q') := \hbar^{-1} 
\e^{\i(pq'-qp')/\hbar}\, (\mathcal{V}_{p'/2,q'/2} - 
\mathcal{V}_{-p'/2,-q'/2})\,.
\end{equation}
In (\ref{fop})  $ p $ and $ q $ play the r\^ole of the 
canonical coordinate functions on phase space and $ \mathcal{V}_{p',q'} 
:= \e^{p'\partial_{p} + q'\partial_{q}} $ is the translation operator
by $ (p',q') $. In other words, $ \mathcal{F}(p',q') $ acts on a 
phase-space function $ f $ according to
\begin{align} 
\bigl( \mathcal{F}(p',q') f \bigr) (p,q) = 
\hbar^{-1}\,\e^{\i(pq'-qp')/\hbar}\,\bigl\{&f(p+p'/2,q+q'/2) \notag \\
& - f(p-p'/2,q-q'/2)\bigr\}\,. 
\end{align}
With these preparations and writing  $ \mathcal{F}_{s}(p',q') :=
\e^{s\mathcal{L}} \mathcal{F}(p',q') \,\e^{-s\mathcal{L}} $ 
for the Dirac picture of $ \mathcal{F}(p',q') $, the averaged dynamics 
can be cast into the form
\begin{equation} \label{ausgang}
\tt = \e^{-t\mathcal{L}} \lim_{\tau \downto 0}\;
\overline{ \Texp\biggl\{ -\i \int_{0}^{t} \!\d s
\intpr \tilde{N}_{s}^{(\tau )}(\d p'\d q')\; \mathcal{F}_{s}(p',q')
\biggr\}}\,.
\end{equation}
To perform the stochastic average in (\ref{ausgang}) we appeal to the 
characteristic functional of Gaussian random variables and conclude that
\begin{align} 
\tt &= \e^{-t\mathcal{L}} \lim_{\tau \downto 0}\,
\Texp\biggl\{ -\frac{1}{2} \int_{0}^{t} \!\d s\int_{0}^{t} \!\d s'
\; g^{(\tau )}(s-s')  \notag\\ 
& \hspace*{3.5cm}\times
\intpr \tilde{C}(\d p'd q')\;
\mathcal{F}_{s}(-p',-q')\;\mathcal{F}_{s'}(p',q')\biggr\}
\notag\\
&= \e^{-t\mathcal{L}} \Texp\biggl\{ -\frac{1}{2} \int_{0}^{t} \!\d s
\intpr \tilde{C}(\d p'd q')\;\mathcal{F}_{s}(-p',-q')\;
\mathcal{F}_{s}(p',q')\biggr\}\,.  \label{zwischenerg} \raisetag{-.6ex}
\end{align}
Furthermore one has
\begin{align}  \label{einklang}
\intpr \tilde{C}(\d p'd q')\;
&\mathcal{F}(-p',-q')\;\mathcal{F}(p',q')           \notag\\
& = \hbar^{-2} \intpr \tilde{C}(\d p'd q')\; 
(2 - \mathcal{V}_{p',q'} - \mathcal{V}_{-p',-q'})   \notag\\
& = 2\,\hbar^{-2}\,\bigl\{ C(0,0) - C(-\i\hbar\partial_{q},
\i\hbar\partial_{p})\bigr\}  \notag\\
& = 2\,\mathcal{N}\,,  
\end{align}
because $ C $ is an even function. Hence
\begin{equation}  \label{rueckkehr}
\tt = \e^{-t\mathcal{L}}\Texp\biggl\{ - \int_{0}^{t}\!\d s\;
\e^{s\mathcal{L}} \mathcal{N} \e^{-s\mathcal{L}} \biggr\} 
= \e^{-t(\mathcal{L}+\mathcal{N})} \,,
\end{equation}
and the proof is complete.
\end{proof}

\begin{bem}
It follows from Statement \ref{tthg} by a differentiation 
with respect to time that the averaged 
state $ \overline{w_{t}} $ obeys the linear integro-differential equation
\begin{align} \label{master}
\partial_{t} \overline{w_{t}}(p,q)  = 
[ \,\overline{w_{t}},  H](p,q)
+  \hbar^{-2} \intpr\tilde{C}(\d p'\d q') \,
\bigl\{&\overline{w_{t}}(p+p',q+q') \notag\\
&- \overline{w_{t}}(p,q)  \bigr\}
\end{align}
with initial condition $ \overline{w_{0}} = w $.
Eq.\ (\ref{master}) is a generalization of the main 
result of \citet{JaKu82}. In the special case of a free deterministic part,
$ H = p^{2}/2m $, and a momentum-independent noise,
that is, when $ C $ does not depend on $ p $, it reduces to an equation
which is equivalent to Eq.\ (8) in \citet{JaKu82}.
The latter is derived there by directly averaging the von Neumann equation
with the help of a partial integration with respect to the Gaussian
probability measure \citep{Fur63,Don64,Nov64}, see also 
\citet[p.\ 32]{RyKrTa89} or \citet[Thm.\ 6.3.1]{GlJa87}.
The present approach is more direct and appears to provide
a better starting point towards a generalization 
to the case of non-vanishing correlation times, that is, coloured noise.
Note also that the subsequent treatment of Eq.\ (8) in \citet{JaKu82} 
is restricted to a pure initial state which, in Weyl-Wigner language, 
is represented by a  Gaussian Wigner density $ w $. 
\end{bem}

Being the stochastic average of a Hamiltonian dynamics the semigroup
$ \tt $ is completely positive and thus provides an example of a
quantum-dynamical semigroup in the sense of \citet{Bau66},
\citet{GoKoSu76}, \citet{Lin76}, \citet{Spo80}, \citet{AlLe87} or
\citet{Str95}.
Quantum-dynamical equations similar to (\ref{master}) are discussed in very 
different branches of physics. One reason for this is revealed by

\begin{satz}  \label{irreversibel}
The averaged dynamics is monotone mixing increasing in the sense that
\begin{equation}\label{mixing}
\partial_{t}\, 
\langle \tt w, \mathcal{T} _{t} w \rangle \le 0 
\end{equation}
for all $ t>0 $ and all Wigner densities $ w $. Here we use the 
notation $ \langle\winzbullet , \winzbullet\rangle $ 
for the standard scalar product \eqref{spur} of phase-space functions. 
In \eqref{mixing} equality holds if and only if $ C $ is a constant. 
\end{satz}

\begin{proof}
Taking the scalar product of (\ref{master}) with
$ \overline{w_{t}} $ and observing
$ \langle \overline{w_{t}}, [ \overline{w_{t}}, H]\rangle = 0 $, which
follows from a partial integration, we get
\begin{equation}  \label{mixbew}
\langle \overline{w_{t}},\partial_{t}\overline{w_{t}}\rangle  = 
\hbar^{-2} \intpr \tilde{C}(\d p'\d q')
\left\{ \langle\overline{w_{t}},\mathcal{V}_{p',q'}\overline{w_{t}}\rangle -
\langle\overline{w_{t}},\overline{w_{t}}\rangle \right\}\,.
\end{equation}
Due to the positivity of the measure $ \tilde{C} $,
inequality (\ref{mixing}) now follows from an application of the
Cauchy-Schwarz inequality
\begin{equation} \label{CSU}
\langle\overline{w_{t}},\mathcal{V}_{p',q'}\overline{w_{t}}\rangle \le
{ \langle\overline{w_{t}},\overline{w_{t}}\rangle }^{1/2}
{ \langle \mathcal{V}_{p',q'}\overline{w_{t}}, 
\mathcal{V}_{p',q'}\overline{w_{t}}\rangle }^{1/2}
= \langle\overline{w_{t}},\overline{w_{t}} \rangle\,.
\end{equation}
To obtain equality in (\ref{mixing}) it is necessary and sufficient that
there is equality in (\ref{CSU}) for $ \tilde{C} $-almost all $ (p',q') $.
Hence, it is necessary and sufficient that the functions
$ \overline{w_{t}} $ and $ \mathcal{V}_{p',q'}\overline{w_{t}} $ 
are linearly dependent for $ \tilde{C} $-almost all $ (p',q') $.
In this case there are constants
$ \lambda_{p',q'} $ with $ \mathcal{V}_{p',q'}
\overline{w_{t}} = \lambda_{p',q'} \overline{w_{t}} $. 
Normalization 
$ \langle 1, \mathcal{V}_{p',q'}\overline{w_{t}}
\rangle = \langle 1, \overline{w_{t}}\rangle $ 
requires 
$ \lambda_{p',q'} = 1 $.
For $ (p',q') \neq (0,0) $ this implies periodicity of the Wigner 
density $ \overline{w_{t}} $ which is in contradiction to its 
square-integrability $ \langle \overline{w_{t}}, \overline{w_{t}}\rangle 
\le (2\pi \hbar)^{-1}<\infty $, cf.\ Statement~\ref{wigprop}. 
To summarize, the Wigner densities $ \overline{w_{t}} $ and 
$ \mathcal{V}_{p',q'}\overline{w_{t}} $ are linearly dependent 
if and only if $ (p',q') = (0,0) $ which in turn allows for equality in
(\ref{mixing}) if and only if $ \tilde{C} $ has support
$ \{(0,0)\} $.
\end{proof}

\noindent
Statement \ref{irreversibel} can be reformulated as
$ \partial_{t} S_{2}(\tt w) \ge 0 $, where 
$ S_{2}(w) := - \ln\bigl( 2\pi \hbar\, \langle w, w\rangle\bigr) $ is the  
R\'enyi entropy of order two of the quantum state \citep{Jum90}.
Similar to the von Neumann entropy, $S_2$ reflects the degree of mixing of the 
quantum state \citep[(2.2,3)]{Thi83} and, by Jensen's inequality, it is a 
non-negative lower bound to the former. 
The average over randomness has turned the 
fully reversible quantum Liouville equation (\ref{liou}) into the ``master
equation'' (\ref{master}) with coherence-destructing irreversible 
behaviour.

So far we have not made use of the fact that the deterministic part 
$ H $ of the Hamiltonian is at most quadratic in $ p $ and $ q $. The next 
result however relies crucially on it.

\begin{satz}       \label{ttww}
The averaged dynamics admits the representation
\begin{equation} \label{ttwweq}
\tt = \e^{-t\mathcal{L}}\;\exp\biggl\{ -\hbar^{-2}\int_{0}^{t}
\!\d s\; \bigl( C(0,0) - C(-\i\hbar\mathcal{K}_{s},
\i\hbar\mathcal{X}_{s})\bigr)\biggr\}\,.
\end{equation}
The coefficients of the first-order differential operators
\begin{equation}\label{megadiffer}
\begin{split} 
\mathcal{K}_{s} & := (\partial_{q}\e^{-s\mathcal{L}}p)\partial_{p}
+ (\partial_{q}\e^{-s\mathcal{L}}q)\partial_{q} \,,\\
\mathcal{X}_{s} & := (\partial_{p}\e^{-s\mathcal{L}}p)\partial_{p}
+ (\partial_{p}\e^{-s\mathcal{L}}q)\partial_{q} 
\end{split}
\end{equation}
depend only on time and not on position and momentum. They are 
determined by the classical phase-space trajectories generated by $ H $.
\end{satz}

\noindent
Note that there is no time ordering in (\ref{ttwweq}) because the four 
operators $ \mathcal{K}_{s}$, $\mathcal{K}_{s'}$, $\mathcal{X}_{s}$,
$\mathcal{X}_{s'} $ commute pairwise due to the quadratic nature of $H$. 
Statement \ref{ttww} is important because it 
allows to derive explicit results on the averaged dynamics $ \tt $.

\begin{proof}[Proof\phantom{x}of\phantom{x}Statement \ref{ttww}]
Thanks to (\ref{rueckkehr}) it only remains to show that
\begin{equation}\label{auffunkt}
\e^{s\mathcal{L}}\mathcal{N}\e^{-s\mathcal{L}} =
\hbar^{-2}\;\bigl( C(0,0) - C(-\i\hbar\mathcal{K}_{s},
\i\hbar\mathcal{X}_{s})\bigr)\,.
\end{equation}
But when applying both sides of (\ref{auffunkt}) to a phase-space function,
this identity follows from 
the chain rule and from observing that the 
phase-space trajectories $ \bigl\{ (e^{t\mathcal{L}}p)(p',q') \bigr. $,
$ \bigl. (e^{t\mathcal{L}}q)(p',q')\bigr\}_{t\ge 0} $ are linear 
functions in the initial data $ (p',q') $ due to the quadratic nature 
of $ H $. 
\end{proof}

\noindent
To gain some more insight into the behaviour of $ \tt $ define
the phase-space function
\begin{equation}     \label{pete}
\tilde{P}_{t}(p,q)  := 
\e^{-t C(0,0)/\hbar^2}\,
\exp\biggl\{ \hbar^{-2} \int_{0}^{t}\!\d s\;
(C{\scriptstyle\raisebox{.05ex}{$\circ$}}\mathsf{J}_{-s}^{\adjoint})(p,q) 
\biggr\}\,,
\end{equation}
where  
\begin{equation}   \label{jacobi}
\mathsf{J}_{t} := 
\begin{pmatrix} 
\partial_{p}\e^{t\mathcal{L}}p & \partial_{q}\e^{t\mathcal{L}}p \\
\partial_{p}\e^{t\mathcal{L}}q & \partial_{q}\e^{t\mathcal{L}}q 
\end{pmatrix}
\end{equation}
is the Jacobian of the flow $ \etl $ on phase space and 
$ \mathsf{J}_{t}^{\adjoint} := 
\bigl(\begin{smallmatrix}0 & -1\\1 & 0\end{smallmatrix}\bigr)
\;\mathsf{J}_{t}^{\,\mathsf{T}}\,
\bigl(\begin{smallmatrix}0 & 1\\-1 & 0\end{smallmatrix}\bigr) $
is its adjoint with respect to the canonical symplectic form 
$ \bigl( (p,q), (p',q')\bigr) \mapsto pq'-qp' $. 
The superscript $ \rule{0pt}{1ex}^{\mathsf{T}} $ denotes the
transposition of matrices and the circle 
${\scriptstyle\raisebox{.05ex}{$\circ$}}$ 
denotes the composition of mappings. Because of 
\begin{equation} 
\begin{pmatrix} 
-\i\hbar\mathcal{K}_{s}\\ 
{}\i\hbar\mathcal{X}_{s}
\end{pmatrix}
= \mathsf{J}_{-s}^{\adjoint}
\begin{pmatrix} 
-\i\hbar\partial_{q}\\ 
{}\i\hbar\partial_{p}
\end{pmatrix}
\end{equation}
one can write $ \tt = \emtl\,\tilde{P}_{t}(-\i\hbar\partial_{q},
\i\hbar\partial_{p}) $. Now observe that $ \tilde{P}_{t} $ is a
characteristic function, that is, the Fourier transform of a measure
$ P_{t} $ on phase space
\begin{equation} 
\tilde{P}_{t}(p,q)  =:  \intpr P_{t}(\d p'\d q')\;
\e^{\i(pq'-qp')/\hbar} \,.
\end{equation}
This is true because 
$ \hbar^{-2}\int_{0}^{t}\!\d s\;  C{\scriptstyle\raisebox{.05ex}{$\circ$}}
\mathsf{J}_{-s}^{\adjoint} $
is a characteristic function and because sums and products of 
characteristic functions are again characteristic functions. 
In fact, $P_{t}$ is even a probability measure due to 
$ \tilde{P}_{t}(0,0) =1 $. Thus we conclude 
\begin{equation} \label{ppositiv}
\tt = \e^{-t\mathcal{L}} \intpr P_{t}(\d p'\d q')\; \mathcal{V}_{p',q'}\,.
\end{equation}

\begin{bem} \label{classinter}
The result (\ref{ppositiv}) shows that the averaged dynamics is also 
positivity preserving and thus even a Markov semigroup. In particular,
it maps classical states, that is, probability densities on $\rxr\,$, to
classical states. Eq.\ (\ref{master})
appears as the associated kinetic equation, which---up to the drift term 
arising from $ H $---may be viewed upon as a linear Boltzmann equation
with a homogeneous stochastic kernel \citep[Chap.\ XI.12]{LaMa94,ReSi79}.
\end{bem}

Since the covariance function $ C $ of the random field is an even
function, all 
odd moments of the probability measure $ P_{t} $ vanish. The covariance 
matrix of $ P_{t} $
\begin{align}   \label{kovmat}
\mathsf{C}_{t} :&= \intpr P_{t}(\d p\d q)
\begin{pmatrix} p^{2}  &  pq  \\  pq  &  q^{2} \end{pmatrix}
= -\hbar^{2} \bigl( (\partial^{2} \tilde{P}_{t})(0,0) 
\bigr)^{\!\adjoint} \notag\\
&= \int_{0}^{t}\!\d s\, \mathsf{J}_{-s}\; 
\bigl((- \partial^{2} C)(0,0)\bigr)^{\!\adjoint}\;
\mathsf{J}_{-s}^{\,\mathsf{T}}
\end{align}
is monotone increasing in time,
\begin{equation}  \label{kovmatwachs}
0\le\mathsf{C}_{t'}\le \mathsf{C}_{t}\,,\qquad 0\le t' \le t\,.
\end{equation}
In (\ref{kovmat}) we have used the symbol $ \partial^2 $ for the Hessian 
of second-order partial derivatives. For later purpose it is
convenient to introduce the non-negative matrix of diffusion coefficients 
\begin{equation} \label{dee}
\mathsf{D} := 
\begin{pmatrix} D_{0,2} & \tfrac{1}{2}\,D_{1,1} \\ 
\tfrac{1}{2}\,D_{1,1} & D_{2,0}\end{pmatrix}
= - \frac{1}{2}\; \bigl((\partial^{2} C)(0,0)\bigr)^{\!\adjoint} \ge 0
\end{equation}
whose entries 
\begin{equation} \label{demunu}
D_{\mu ,\nu }:= \bigl( (-i\partial_{p})^{\mu }
(i\partial_{q})^{\nu } C\bigr) (0,0) / (\mu !\, \nu !)
\end{equation}
are given by the curvature of the covariance function at the
origin. Thus one can also write 
\begin{equation}
\mathsf{C}_{t} = 2 \int_{0}^{t}\!\d s\, \mathsf{J}_{-s}\;\mathsf{D}\;
\mathsf{J}_{-s}^{\,\mathsf{T}}\,.  
\end{equation}

It is clear from the Riemann-Lebesgue lemma that the probability measure
$ P_{t} $ cannot be absolutely continuous with respect to Lebesgue measure
$ \d p\d q $. The following decomposition helps to decide whether $ P_{t} $
possesses an absolutely continuous component.

\begin{satz}\label{ttzer}
The averaged dynamics of an initial state $ w $ 
\begin{align}\label{zerlegung}
(\tt w)(p,q)  =& \;  \e^{-tC(0,0)/\hbar^{2}}\;
w(\e^{-t\mathcal{L}}p,\e^{-t\mathcal{L}}q)   \notag\\  
&+ \intpr Q_{t}(\d p'\d q')\; w\bigl(p' + \e^{-t\mathcal{L}}p,
q' + \e^{-t\mathcal{L}}q\bigr)
\end{align}
decomposes into an exponentially decaying contribution of the 
unperturbed dynamics and into a contribution where the initial state 
is first smeared out by the finite measure $ Q_{t} $ and then subjected 
to the unperturbed dynamics. The measure $ Q_{t} $ 
is defined by its characteristic function
\begin{equation} \label{qute}
\tilde{Q}_{t} := \tilde{P}_{t} - \e^{-t C(0,0)/\hbar^2}
\end{equation}
and is normalized to $ \tilde{Q}_{t}(0,0)= 1- \e^{-t C(0,0)/\hbar^2} $.
All higher moments of $ Q_{t} $ coincide with those of $ P_{t} $.
A sufficient condition for $ Q_{t} $ to have an integrable and 
continuous density with respect to Lebesgue measure is the integrability 
of $ C $. The density is then bounded by
\begin{equation}   \label{tauschranken}
0 \le \frac{Q_{t}(\d p\d q)}{\d p\d q}\le \frac{t}{(2\pi \hbar^{2})^2} 
\intpr\d p'\d q'\; |C(p',q')| \,.
\end{equation}
Necessary conditions for the existence of a density are the strict 
positivity of the covariance matrix, $ \mathsf{C}_{t} > 0 $, 
and the vanishing of the function
$ (p,q)\mapsto\int_{0}^{t}\!\d s\, (C{\scriptstyle\raisebox{.05ex}{$\circ$}} 
\mathsf{J}_{-s}^{\adjoint})(p,q) $
at infinity.
\end{satz}

\begin{figure} 
\newcommand{\gauss}{
  \scaleput(0,0){
     \qbezier(-.25,1)(-.15,3)(0,3)
     \qbezier(.25,1)(.15,3)(0,3)    
     \qbezier(.25,1)(.275,.5)(1,.3)
     \qbezier(-.25,1)(-.275,.5)(-1,.3)
  }
}
\newcommand{\breitgauss}{
  \scaleput(0,0){
     \curvedashes[.25mm]{2.6,2.8}
     \qbezier(-.5,1)(-.3,2)(0,2)
     \qbezier(.5,1)(.3,2)(0,2)    
     \qbezier(1,.3)(.6,.5)(.5,1)
     \qbezier(-1,.3)(-.6,.5)(-.5,1)
  }
}
\newcommand{\bbreitgauss}{
  \scaleput(0,0){
     \curvedashes[.4mm]{2.4,2.6}
     \qbezier(0,2)(-.2,2)(-.5,1)
     \qbezier(0,2)(.2,2)(.5,1)    
     \qbezier(1,.3)(.75,.4)(.5,1)
     \qbezier(-1,.3)(-.75,.4)(-.5,1)
  }
}

\unitlength.8cm
\begin{picture}(13.7,11)
\thinlines
\put(.6,4.8){\vector(1,0){12.3}}                  
\put(12.8,4.4){\small $ q $}
\put(6.4,4.8){\vector(0,1){5.9}}                  
\put(6.65,10.4){\small $ \mathcal{T}_{t} w $}
{\curvedashes[.5mm]{1.85,1}\put(6.4,4.8){\curve(0,-1, 0,0)}}
\renewcommand{\yscalex}{-.55}                     
\renewcommand{\xscaley}{.55}
\renewcommand{\yscale}{.85}
\renewcommand{\xscale}{.85}
\put(6.4,4.8){\scaleput(0,0){\curve(0,-4.7, 0,5.7)}
              \scaleput(-0.043,5.7){\begin{turn}{-6}\vector(1,2){0}\end{turn}}}
\put(9.8,9.4){\small $ p $}  
\put(6.65,4.8){
  \begin{picture}(12.5,9.5)
  \linethickness{.75pt}
  \renewcommand{\yscalex}{0}
  \renewcommand{\xscaley}{0}
  \renewcommand{\yscale}{1}
  \renewcommand{\xscale}{1.2}
  \put(3.5,-.2){\gauss}
  \put(4.2,1.8){\small $ t=0 $}
  \renewcommand{\xscale}{1}
  \renewcommand{\yscale}{.83}
  \put(1,-3.4){\gauss}
  \renewcommand{\xscale}{.6}
  \renewcommand{\yscale}{.25}
  \put(1,-3.22){\breitgauss}
  \put(2.8,-3.2){\small $ t=t_{1} > 0 $}
  \renewcommand{\xscale}{.3}
  \renewcommand{\yscale}{.25}
  \put(-3.1,.9){\gauss}
  \renewcommand{\xscale}{1.8}
  \renewcommand{\yscale}{.55}
  \put(-3.1,.8){\bbreitgauss}
  \put(-5.5,2.5){\small $ t=t_{2} > t_{1} $}
  
%
%
  \linethickness{.5pt}
  \renewcommand{\xscale}{.71}          
  \renewcommand{\xscaley}{-.55}
  \renewcommand{\yscalex}{.71}
  \renewcommand{\yscale}{.55}
  \scaleput(0,0){\arc[8](4,0){54.5}}
  \scaleput(0,0){\arc[3](4,0){-8}}
  \scaleput(0,0){\arc[6](0,4){-28}}
  \scaleput(0,0){\arc[2](0,4){5}}
  \scaleput(0,0){\arc[9](-4,0){66}}
  \scaleput(0,0){\arc[9](-4,0){-65}}
  \scaleput(0,0){\arc[10](0,-4){40}}  
  \scaleput(0,0){\arc[5](0,-4){-11}}  
%
%
  \linethickness{.5pt}
  \renewcommand{\xscale}{-.61}
  \renewcommand{\yscale}{-.61}
  \renewcommand{\xscaley}{.81}
  \renewcommand{\yscalex}{-.81}
  \put(2.57,-1.8){\scaleput(0,0){\curve(0,0, -.18,-.08)}}
  \put(2.57,-1.8){\scaleput(0,0){\curve(0,0, -.18,.08)}}
  \renewcommand{\xscale}{0}
  \renewcommand{\yscale}{0}
  \renewcommand{\xscaley}{1}
  \renewcommand{\yscalex}{-1}
  \put(-3.59,-.97){\scaleput(0,0){\curve(0,0, .18,-.08)}}
  \put(-3.59,-.97){\scaleput(0,0){\curve(0,0, .18,.08)}}
  \end{picture}
}
\end{picture}
\caption{Illustration of the averaged dynamics of a Wigner density in phase 
space. The deterministic part of the Hamiltonian has been chosen to be
that of a harmonic oscillator. The solid (dashed) contribution
symbolizes the first (second) term in the decomposition 
(\protect{\ref{zerlegung}}).
}%

\label{geometrie}
\end{figure}
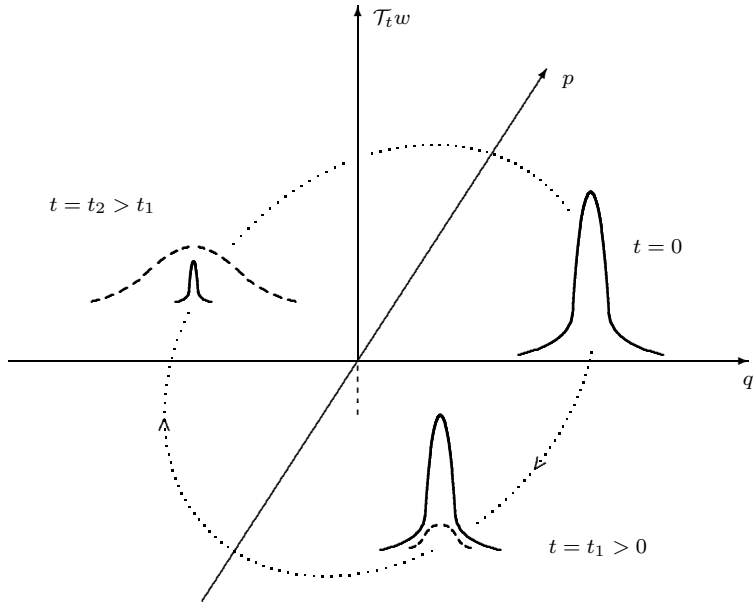

\noindent
Figure \ref{geometrie} illustrates the situation described in Statement
\ref{ttzer} for the case where $ H $ is the Hamiltonian of a 
harmonic oscillator and $ w $ a non-negative initial Wigner density at 
time $ t=0 $. 
The dotted ellipse represents one phase-space trajectory 
of $ H $. The averaged state $ \tt w $ at times $ t_{1} > 0 $ and
$ t_{2}> t_{1} $
is obtained by summing up the solid and the dashed contribution which
symbolize the first and second term in the decomposition 
(\ref{zerlegung}), respectively. Accordingly, the weight of the 
solid contribution to $ \tt w $ decreases exponentially in time, 
whereas that of the dashed one increases to one. 
In addition, the width of the dashed contribution increases in time.
The sum of the volumes enclosed underneath is always one.

\begin{proof}[Proof\phantom{x}of\phantom{x}Statement \ref{ttzer}]
The arguments which established that $ \tilde{P}_{t} $ is a 
characteristic function apply to $ \tilde{Q}_{t} $ as well. Hence,
$ Q_{t} $ is a finite measure on $ \rxr\, $. According to
\citet[{}\S~23, Problem~10]{Bau96} the integrability of $ \tilde{Q}_{t} $
is sufficient to guarantee the existence and continuity of the
Lebesgue density
$ Q_{t}(\d p\d q)/\d p\d q $. But the inequality $ |\e^{z} -1| \le 
|z| \e^{y} $, which is valid for all $ |z|\le y $, shows that this 
condition is fulfilled if $ C $ is integrable. Here, we have also used 
the fact that phase-space volume is conserved by Hamiltonian flows.
The upper bound in (\ref{tauschranken}) then follows from the Fourier 
inversion formula. The second of the two necessary conditions is just 
the content of the Riemann-Lebesgue lemma. As to the first one we remark 
that if it did not hold, there would exist a pair $ (p',q') $
such that $ \intpr Q_{t}(\d p\d q)\; (p'q -q'p)^{2} = 0 $. 
But this implies that $ Q_{t} $ is supported on a set of Lebesgue 
measure zero.
\end{proof}

\subsection{The classical limit} \label{class}

One main advantage of a phase-space representation of quantum
mechanics is that the classical limit of a quantum system may be
studied with little efforts. We denote the averaged dynamics of the
classical white-noise system by
\begin{equation} 
  \ttkl := \lim_{\hbar\downto 0} \tt
\end{equation}
and conclude

\begin{satz} \label{ttkl}
  Suppose that the deterministic part $H$ of the Hamiltonian and the
  covariance function $C$ do not depend on $\hbar$.\\
  Then the averaged dynamics of the classical white-noise system 
  \begin{enumerate}
  \item is a semigroup of linear operators 
    \begin{equation} \label{ttklhg}
      \ttkl = \exp\{-t (\mathcal{L} + \mathcal{N}_{\mathrm{cl}})\}
      = \e^{-t \mathcal{L}}\;\exp\biggl\{ \frac{1}{2}\,
      (\partial_{p},\partial_{q})\, \mathsf{C}_{t}\,
      \binom{\partial_{p}}{\partial_{q}}\biggr\}
    \end{equation}
    with generator $\mathcal{L} + \mathcal{N}_{\mathrm{cl}}\,$, where 
    \begin{equation}  \label{enkl}
      \mathcal{N}_{\mathrm{cl}} := -(\partial_{p},\partial_{q}) \,\mathsf{D}
      \,\binom{\partial_{p}}{\partial_{q}}\,.
    \end{equation}
    The covariance matrix $\mathsf{C}_{t}$ and the matrix $\mathsf{D}$
    of diffusion coefficients were introduced in \eqref{kovmat} and
    \eqref{dee}, respectively.
  \item maps an initial probability density $\rho$ on phase space
    to the probability density $\overline{\rho_{t}}:= \ttkl\rho$
    which obeys the Fokker-Planck equation
    \begin{equation} \label{fokker}
      \partial_{t} \overline{\rho_{t}}  = 
      \bigl( (\partial_{q} H) \partial_{p} - (\partial_{p} H) \partial_{q}
      \bigr) \overline{\rho_{t}}  +   (\partial_{p},\partial_{q})
      \,\mathsf{D}\, 
      \binom{\partial_{p}}{\partial_{q}} \overline{\rho_{t}}
    \end{equation}
    with initial condition $\overline{\rho_{0}}=\rho$. The solution of
    \eqref{fokker} is obtained by smearing out $\rho$ with
    the Gaussian probability measure $P_{t,\mathrm{cl}}$ on phase
    space, which is defined by its characteristic function 
    \begin{equation}
      \exp\biggl\{ - \frac{1}{2}\, (x,k)\,\mathsf{C}_{t}\,\binom{x}{k}\biggr\}
      =: \intpr P_{t,\mathrm{cl}} (\d p'\d q')\; \e^{-\i (xp'+kq')}\,,
    \end{equation}
    and subjecting the result to the time evolution of $H$
    \begin{equation}   \label{ptkl}
      \overline{\rho_{t}}(p,q) =  \intpr P_{t,\mathrm{cl}} (\d p'\d
      q')\; \rho 
      (p'+ \e^{-t\mathcal{L}}p, q'+ \e^{-t\mathcal{L}}q)\;.
    \end{equation}
    The measure $P_{t,\mathrm{cl}}$ has zero mean and $\mathsf{C}_{t}$
    as its covariance matrix. It possesses the Lebesgue density 
    \begin{equation}
      \frac{P_{t,\mathrm{cl}}(\d p\d q)}{\d p\d q} = \bigl( \det (2\pi 
      \mathsf{C}_{t})\bigr)^{-1/2}\;\exp\biggl\{ - \frac{1}{2}\, (p,q) \,
      \mathsf{C}_{t}^{-1}\, \binom{p}{q}\biggr\} 
    \end{equation}
    if and only if $\mathsf{C}_{t}>0$.
  \item is irreversible, if and only if the covariance function 
    $C$ is non-constant. In this
    case the Boltzmann-Gibbs entropy 
    \begin{equation}
      S(\overline{\rho_{t}}) := - \bigl\langle \overline{\rho_{t}} ,
      \ln(2\pi\hbar\overline{\rho_{t}})\bigr\rangle 
    \end{equation}
    of $\overline{\rho_{t}}$ (if existent) is a strictly increasing
    function in time. 
  \end{enumerate}
\end{satz}

\begin{proof}
  The first part of Statement \ref{ttkl} follows from Taylor expanding
  $\mathcal{N}$ in \eqref{rueckkehr} with respect to $\hbar$
  \begin{equation} \label{enexpand}
    \mathcal{N} = -D_{0,2}\partial_{p}^2 -D_{1,1}\partial_{p}\partial_{q}
    -D_{2,0}\partial_{q}^2 + \mathcal{O}(\hbar^2 \partial^4)\,,
  \end{equation}
  with $D_{\mu ,\nu}$ being defined in \eqref{demunu}. In the limit
  $\hbar\downto 0$ the expansion \eqref{enexpand} reduces to the semielliptic
  differential operator $\mathcal{N}_{\mathrm{cl}}$. To obtain the
  right equality in \eqref{ttklhg} it is useful to note in addition
  the operator identity 
  \begin{equation} \label{ophelpid}
    \binom{\partial_{p}}{\partial_{q}}\e^{-s \mathcal{L}} =
    \e^{-s \mathcal{L}} \mathsf{J}_{-s}^{\,\mathsf{T}}
    \binom{\partial_{p}}{\partial_{q}}\,,
  \end{equation}
  which is a consequence of the chain rule and may be proved by
  applying both sides of \eqref{ophelpid} to a phase-space function.
  The second part of Statement \ref{ttkl} is an immediate consequence
  of the first part. To prove the third part, suppose that the
  Boltzmann-Gibbs entropy $S(\ttkl\rho)$ exists. Since $\ttkl$ is a
  semigroup it suffices to show $S(\ttkl\rho)>S(\rho)$ if $t>0$ and
  $C$ is non-constant. To do so we conclude from \eqref{ptkl}
  \begin{align} \label{bgentropy}
    S(\ttkl\rho) &= - \int_{\rxr}\! P_{t,\mathrm{cl}}(\d p'\d q')\;
    \Bigl\langle \rho , \ln \bigl(2\pi\hbar
    \mathcal{V}_{-p',-q'}\etl\ttkl\rho \bigr) \Bigr\rangle \notag\\
    &> - \int_{\rxr}\! P_{t,\mathrm{cl}}(\d p'\d q')\;\bigl\langle
    \rho , \ln (2\pi\hbar\rho ) \bigr\rangle \notag\\ 
    &= S(\rho)\,,
  \end{align}
  where $\mathcal{V}_{p',q'}=\e^{p'\partial_{p} + q'\partial_{q}}$ denotes
  the translation operator by $(p',q')$. The strict inequality in
  \eqref{bgentropy} follows from the elementary inequality 
  \begin{equation}
    u - u \ln u < v - u \ln v\,,\qquad\quad u,v\ge0\,, u\neq v\,.
  \end{equation}
  To see this we note that for $t>0$ and a non-constant $C$, the support
  of $P_{t,\mathrm{cl}}$ is at least a one-dimensional subspace of $\rxr\,$.
  The integrability of $\rho$ therefore forbids the equality 
  \begin{equation}
    \rho (p,q) = \int_{\rxr}\! P_{t,\mathrm{cl}}(\d p''\d q'')\;
    \rho (p''-p'+p,q''-q'+q)
  \end{equation}
  for $\rho\d p\d q$-almost all $(p,q)\in\rxr$ and 
  $P_{t,\mathrm{cl}}(\d p'\d q')$-almost all $(p',q')\in\rxr\,$.  
\end{proof}

\begin{bem}
  The classical phase-space trajectories $(p_{t},q_{t})$ generated by
  the Hamiltonian $H+N_{t}$ realize a diffusion process on phase
  space. Its underlying stochastic differential equation coincides
  with Hamilton's equations of motion
  \begin{equation}
    \begin{split} \label{langevin}
      \partial_{t} p_{t} & = -\bigl(\partial_{q}( H + N_{t})\bigr)
      (p_{t},q_{t})\,,\\ 
      \partial_{t} q_{t} & =  \bigl(\partial_{p}( H + N_{t})\bigr)
      (p_{t},q_{t}) \,,
    \end{split}
  \end{equation}
  and \eqref{fokker} is the associated Fokker-Planck equation.
\end{bem}

\subsection{Explicit results for averaged expectation values} \label{val}

Now we return to the quantum-mechanical situation.
In order to compute the averaged expectation value of an observable $ a $, 
it is often advantageous to switch to the Heisenberg picture according 
to
\begin{equation}
\overline{\langle w_{t}, a \rangle} =
\langle\overline{ w_{t}}, a \rangle =
\langle\tt w , a\rangle = \langle w, \ttad a\rangle \,.
\end{equation}
The thus defined adjoint semigroup $ \ttad $ is obtained from $ \tt $ by 
simply reversing the sign of $ \mathcal{L} $.   
In analogy to Statement \ref{ttww} the adjoint semigroup has the 
representation
\begin{subequations}            
\begin{align} 
\ttad  &= \e^{t\mathcal{L}}\;
\exp\biggl\{ -\int_{0}^{t}\!\d s\; \e^{-s\mathcal{L}}\;
\mathcal{N}\;\e^{s\mathcal{L}} \biggr\}\label{ttadwwdar0}\\ 
\label{ttadwwdar}
&= \e^{t\mathcal{L}}\;
\exp\biggl\{ -\hbar^{-2}\int_{0}^{t}
\!\d s\; \bigl( C(0,0) - C(-\i\hbar\mathcal{K}_{-s},
\i\hbar\mathcal{X}_{-s})\bigr)\biggr\}\,.
\end{align}
\end{subequations}            
Statement \ref{ttzer} finds a similar translation.
The expression (\ref{ttadwwdar}) serves as our starting point for obtaining 
some explicit results on the averaged dynamics. For example, the averaged 
Heisenberg picture of any observable which is a polynomial in $ p $ and 
$ q $ may be calculated from it. To this end we remark that there are 
only finitely many terms in the Taylor expansions of the right 
exponential in (\ref{ttadwwdar}) and of the covariance function which 
contribute to the result. Examples for the cases where the deterministic 
part of the Hamiltonian is that of a free particle or that of a particle 
subjected to a constant magnetic field in two space dimensions have 
been reported in \citet{FiLeMu94}. Here we only recall the free-particle 
case.

\begin{bsp}   \label{frei}
Let $ H= p^2/(2m) $. Then 
\begin{subequations}
\begin{gather}
\ttad p = p \,,\\
\ttad q = q+ tp/m \,,\\
\ttad p^2 = p^2 + 2 t D_{0,2} \,,
\label{freip2}\\
\ttad pq  = p (q +tp/m)  + t D_{1,1} + t^2 D_{0,2}/m \,, 
\label{freipq}\\
\ttad q^2 = (q + tp/m)^{2} + 2 t D_{2,0} + t^2 D_{1,1}/m
+ \tfrac{2}{3}\, t^3 D_{0,2}/m^2  \,,
\label{freiq2}\\
\ttad q^4 = (q + tp/m)^4 + 
12 \bigl\{ (q + tp/m)^2  d_{2}(t) +
2\hbar^2 d_{4}(t) + \bigl( d_{2}(t)\bigr)^2\bigr\}\label{freiq4}\,.
\end{gather}
\end{subequations}
The coefficients  $ D_{\mu ,\nu } $ were defined in
\eqref{demunu} and we have introduced the polynomials $ \displaystyle
d_{\mu }(t) := \sum_{\nu =0}^\mu \frac{D_{\mu -\nu ,\nu }}{(\nu +1)\,
m^\nu }\; t^{\nu +1}$ of maximal degree $\mu +1$.
As a consequence, the long-time spreading of states under the 
averaged dynamics is diffusive in momentum and superballistic in position
\begin{subequations}
\begin{gather} 
\lim_{t\to\infty } \frac{1}{t}\,\Bigl(
\langle \overline{w_{t}},p^2\rangle - \langle \overline{w_{t}},p\rangle^2
\Bigr) =  2 D_{0,2}\,,\\
\lim_{t\to\infty } \frac{1}{t^3}\,\Bigl(
\langle \overline{w_{t}},q^2\rangle - \langle \overline{w_{t}},q\rangle^2
\Bigr) =  \frac{2 D_{0,2}}{3m^2}\,.
\end{gather}
\end{subequations}
\end{bsp}

Example \ref{frei} illustrates important features valid for
general quadratic $ H $: Observables which are linear
in $ p $ and $ q $ are not affected by the white noise
$ N_{t} $. Moreover, noise-induced terms in averaged expectation values
of quadratic observables are independent of the 
initial state. Assuming that the deterministic part $H$ of the
Hamiltonian  and the covariance function $C$ are $ \hbar $-independent,
noise-induced effects affected by quantum fluctuations occur  
only for observables of at least fourth order in $ p $ or
$ q $. However, very special situations are needed for quantum effects 
to show up in the leading term for long times.
For example, taking the observables $ p^{n} $ or $ q^{n},\; 4\le n
~\mathrm{integer} $, one must 
require the phase-space trajectories of $ H $ to grow exponentially in 
time. These results generalize some earlier ones in 
\citet{JaKu82}, \citet{Hei83b}, \citet{GhRiWe86}, \citet{Hei92} and
\citet{Jay93}.

Due to the time dependence of the random part $ N_{t} $  of the 
Hamiltonian, one does not expect energy conservation for white-noise 
systems. The observable $ \ttad H $ can be interpreted as the mean 
energy of the system at time $ t $. For the special case considered in 
Example \ref{frei}, the mean energy grows linearly in time, see
Eq.\ (\ref{freip2}).
This is also true under more general circumstances.

\begin{satz}\label{tth}
For an at most quadratic deterministic part $H$, the mean energy 
changes at a constant rate:
\begin{equation}
\ttad H = H + t\,\bigl( D_{0,2} (\partial_{p}^2 H) + D_{1,1} (\partial_{p}
\partial_{q} H) + D_{2,0} (\partial_{q}^2 H)\bigr)\,.
\end{equation}
The rate does not allow for a dependence on the initial state.
\end{satz}

\begin{proof}
The claim follows from (\ref{ttadwwdar0}), from $ \e^{s\mathcal{L}} H=H $, 
from
\begin{equation}\label{nh}
-\mathcal{N} H = D_{0,2} (\partial_{p}^2 H) + D_{1,1} (\partial_{p}
\partial_{q} H) + D_{2,0} (\partial_{q}^2 H)
\end{equation}
and from $ \mathcal{N}^2 H = 0 $.
Note also that the right-hand side in (\ref{nh}) does not depend on
$ p $ and $ q $.
\end{proof}

\noindent
Unless $ H $ is unbounded from below, such as $ H= (p^2 - D q^2)/(2m) $  
for $ D> D_{0,2}/D_{2,0} $, the mean energy increases to infinity in 
the course of time. To compensate 
for this effect and, if possible, to allow for an eventual approach to a 
stationary equilibrium state, dissipation has to be incorporated. This 
will be done in the next section by coupling the white-noise system to a 
heat bath.


\section{White noise and dissipation} \label{diss}

This section is devoted to the interplay of white noise and dissipation.
Such an investigation is physically sensible, because the additive
white noise in Model \ref{weissrausch} is in general not due to
thermal fluctuations.
We will show that the incorporation of dissipation causes the mean
energy of the white-noise system to saturate for long times. Topics of this 
kind were already addressed to in \citet{Hei83a}, where dissipation
is generated through an \emph{ad-hoc} time dependence of the
deterministic part of the Hamiltonian.
In contrast, we introduce dissipation by letting the white-noise system 
interact with a microscopic model of a heat bath---a more physical
strategy which is commonly accepted today
\citep{Mag59,FeVe63,FoKaMa65,Ull66,Ara81,CaLe83,HaRe85,FoLeOC88a,GrScIn88,%
SoFu90,Wei93,EfvW94,JaPi97a}.

The programme of this section will be accomplished in four steps:
First, the white-noise system is bilinearly coupled to a heat bath 
of independent harmonic oscillators. The averaged 
dynamics of the total system, that is, of the white-noise system plus
heat bath, is then studied 
with the methods of the preceding section. Second, the bath variables 
are eliminated assuming that the bath is in thermal
equilibrium. Third, the thus obtained reduced averaged dynamics is considered 
in the macroscopic limit of the heat bath. Finally, the long-time
limit is analyzed. Statement \ref{klimax} summarizes the main result.

\subsection{The averaged dynamics of the total system} \label{avtot}

For notational simplicity, we will no longer allow for a momentum
dependence of the  
white noise and we will restrict the deterministic part of the Hamiltonian
to that of a harmonic oscillator. Note also that 
in this section we are using the Weyl-Wigner-Moyal representation for a 
quantum-mechanical system with several Cartesian degrees of freedom.

\begin{model}\label{raureidef} 
The Hamiltonian of the total system is defined by the Weyl-Wigner symbol
\begin{equation}\label{raureibuntham}
H^{(n)} + N_{t}
\end{equation}
on phase space $ \rz^{n+1}\times\rz^{n+1} $.
The zeroth components $(p,q)$ of  
$ \;\mathbf{p}:= (p,p_{1},\ldots ,$ $p_{n}) \in \rz^{n+1} $ and 
$ \mathbf{q}:= (q,q_{1},\ldots , q_{n}) \in \rz^{n+1} $  make up the 
phase-space coordinates of the white-noise system, the following 
components those of the heat bath.
The deterministic part
\begin{equation}\label{raureidetham}
H^{(n)}(\mathbf{p},\mathbf{q}) :=
p^2/(2m) +  (m\omega ^2/2) q^2  +
\sum_{j=1}^n \left\{ p_{j}^2/(2m_{j}) + (m_{j}\omega _{j}^2/2) 
(q_{j} - q)^2 \right\}
\end{equation}
of \eqref{raureibuntham} couples a harmonic oscillator of mass 
$ m>0 $ and frequency $ \omega \ge 0 $ bilinearly to a heat bath of $ n $
independent harmonic oscillators with masses 
$ m_{j}>0 $, $ j=1,\ldots,n $, and frequencies
$ 0<\omega _{1}<\omega _{2}< \ldots < \omega _{n} $. The random part
\begin{equation} \label{raureizufham}
N_{t}(\mathbf{p},\mathbf{q})\equiv N_{t}(q) 
\end{equation}
defines a time-dependent local potential on the phase space of the 
white-noise system only. It is a Gaussian random field
with zero mean and covariance function
\begin{equation} \label{raureikov}
\overline{N_{t}(q)\,N_{t'}(q')} = \delta (t-t')\, C(q-q')\,.
\end{equation}
\end{model}

\vspace{1ex}
Introducing the diagonal $ (n+1)\times(n+1) $-mass matrix
\begin{equation}\label{emmat}
\mathsf{M}:= \mathrm{diag}(m,m_{1},m_{2},\ldots,m_{n}) >0 
\end{equation}
and the $ (n+1)\times(n+1) $-matrix of squared frequencies
\begin{equation}\label{ommat}
\Omega ^2 := \begin{pmatrix} \omega ^2 + \sum\limits_{j=1}^n \kappa _{j}^2
\omega _{j}^2 & -\kappa _{1}\omega _{1}^2 & \phantom{x} & 
\cdot\;\cdot\;\cdot &\phantom{x} & -\kappa _{n}\omega _{n}^2 \\[2ex]
-\kappa _{1}\omega _{1}^2 & \omega _{1}^2 &  &  &  & 0\\
\cdot &  & \cdot \\ 
\cdot &  &  & \cdot \\ 
\cdot &  &  &  & \cdot \\
-\kappa _{n}\omega _{n}^2 & 0 &  &  &  & \omega _{n}^2 
\end{pmatrix} \ge 0 \,,
\end{equation}
where $ \kappa _{j}:= \sqrt{m_{j}/m} $, $ j=1,\ldots,n $, one can rewrite
(\ref{raureidetham}) as a non-negative quadratic form
\begin{equation}\label{quadform}
H^{(n)}(\mathbf{p},\mathbf{q}) = \tfrac{1}{2}\, 
\mathbf{p}\cdot\mathsf{M}^{-1}\mathbf{p} +  \tfrac{1}{2}\,
\mathbf{q}\cdot\mathsf{M}^{1/2}\Omega ^2\mathsf{M}^{1/2}\mathbf{q}\,.
\end{equation}
This model for the heat bath is referred to in the literature as the 
independent oscillator model \citep{FoLeOC88a,FoLeOC88b}
or Caldeira-Leggett model \citep{Wei93}.

Following the strategy of the preceding section we define 
the averaged dynamics of the total system as
\begin{equation}   \label{gemgesdyn} 
\ttn := \lim_{\tau \downto 0}\;\overline{
\Texp \left\{ - \int_{0}^t\!\d s\; 
[ H^{(n)} + N_{s}^{(\tau )}, \winzbullet ] \right\}
}\,,
\end{equation}
where $ N_{t}^{(\tau)} $ is a Gaussian coloured noise which converges to 
the Gaussian white noise $ N_{t} $ of (\ref{raureizufham}) in the limit 
of vanishing correlation time $ \tau \downto 0 $, confer 
Subsection~\ref{wnham}.

The validity of the following statement depends only on the quadratic 
form (\ref{quadform}) of $ H^{(n)} $. The specific forms of the matrices 
$ \mathsf{M} $ and $ \Omega ^2 $ are irrelevant.

\begin{satz}\label{raureiww}
The averaged dynamics $ \ttn $ is a semigroup and admits the 
representation
\begin{equation}\label{raureiwwglg}
\ttn = \e^{-t\mathcal{L}^{(n)}}\;\exp\biggl\{ -\hbar^{-2}\int_{0}^{t}
\!\d s\; \bigl( C(0) - C(\i\hbar\mathcal{X}_{s}^{(n)})\bigr)\biggr\}\,.
\end{equation}
Here we have introduced the differential operator
\begin{equation}  \label{xsn}
\mathcal{X}_{s}^{(n)}:= \left( \mathsf{M}^{-1/2}\cos (\Omega s)
\mathsf{M}^{1/2}\,\partial_{\mathbf{p}}  -
\mathsf{M}^{-1/2}\;\frac{\sin (\Omega s)}{\Omega}\;\mathsf{M}^{-1/2}\,
\partial_{\mathbf{q}} \right)_{0}
\end{equation}
as the zeroth component of the vector on the right-hand side of
\eqref{xsn}. The superscript $n$ accounts for the fact that the Liouvillian 
which generates the reversible dynamics of the deterministic part of the 
total Hamiltonian is $ \mathcal{L}^{(n)}:= [ H^{(n)},\winzbullet ] $.
\end{satz}

\begin{proof}
The semigroup property is established as in the proof of Statement
\ref{tthg}. To arrive at the representation (\ref{raureiwwglg}) 
we remark that the classical phase-space trajectories generated by 
$ \mathcal{L}^{(n)} $ are given by
\begin{equation}
\begin{split}
\etln\mathbf{p}  &=  \mathsf{M}^{1/2}\cos (\Omega t) \mathsf{M}^{-1/2}
\,\mathbf{p} -   \mathsf{M}^{1/2}\Omega\sin (\Omega t) \mathsf{M}^{1/2}\,
\mathbf{q}\,,\\
\etln\mathbf{q}  &=  \mathsf{M}^{-1/2}\;\frac{\sin (\Omega t)}{\Omega }\; 
\mathsf{M}^{-1/2} \,\mathbf{p} + 
\mathsf{M}^{-1/2}\cos (\Omega t) \mathsf{M}^{1/2}\,\mathbf{q}
\end{split}
\end{equation}
and that the differential operators $ \mathcal{K}_{s} $ and 
$ \mathcal{X}_{s} $ introduced in Statement \ref{ttww} are to be 
replaced in the multi-dimensional case by vector-valued differential 
operators whose $ j $-th components, $j\in\{ 0,1,\ldots ,n\}$, read
\begin{subequations}\label{vecdiff}
\begin{align}
(\pmb{\mathcal{K}}_{s})_{j} & := (\partial_{q_{j}}\e^{-s\mathcal{L}}
\mathbf{p})\cdot\partial_{\mathbf{p}}
+ (\partial_{q_{j}}\e^{-s\mathcal{L}}\mathbf{q})\cdot
\partial_{\mathbf{q}} \,, \label{vecdiffk} \\ 
(\pmb{\mathcal{X}}_{s})_{j} & := (\partial_{p_{j}}\e^{-s\mathcal{L}}
\mathbf{p})\cdot\partial_{\mathbf{p}}
+ (\partial_{p_{j}}\e^{-s\mathcal{L}}\mathbf{q})\cdot
\partial_{\mathbf{q}}\,.  \label{vecdiffx}
\end{align}
\end{subequations}
In this notation Definition (\ref{xsn}) appears as 
$ \mathcal{X}_{s}^{(n)} = (\pmb{\mathcal{X}}_{s})_{0} $.
\end{proof}

\noindent
To describe the averaged dynamics of observables we introduce the 
adjoint $ \ttnad $ of the semigroup $ \ttn $ with respect to the 
standard scalar product
\begin{equation} \label{nskalar}
\langle f, g\rangle := \int_{\rz^{n+1}\times\rz^{n+1}}\!
\d^{n+1}p\,\d^{n+1}q\; f^{*}(\mathbf{p},\mathbf{q}) 
g(\mathbf{p},\mathbf{q})
\end{equation}
for square-integrable functions $ f,g $ on phase space
$ \rz^{n+1}\times\rz^{n+1} $. 

Since we are mostly interested in the 
behaviour of the mean energy, it suffices to concentrate on the 
averaged dynamics of the system observables $ p^2 $ and $ q^2 $,
and thus on the spreading of an averaged state of the white-noise system. 
Again, explicit results on the averaged 
dynamics of polynomial observables may be obtained from a Taylor 
expansion of the adjoint semigroup.

\begin{satz} \label{zuraureiww}
\begin{subequations}\label{raurei2}
\begin{align} 
\label{raureip}
\ttnad p^2 &= \bigl( \etln p\bigr)^2 + (-\partial_{q}^2 C)(0)
\int_{0}^t\!\d s\; \left\{ \bigl( \mathsf{M}^{-1/2}\cos (\Omega s)
\mathsf{M}^{1/2}\bigr)_{00}\right\}^2 \,, \\
\label{raureiq}
\ttnad q^2 &= \bigl( \etln q\bigr)^2 + (-\partial_{q}^2 C)(0)
\int_{0}^t\!\d s\; \left\{ \Bigl( \mathsf{M}^{-1/2}
\frac{\sin (\Omega s)}{\Omega }\,\mathsf{M}^{-1/2}\Bigr)_{00}\right\}^2 .
\raisetag{-1ex}
\end{align}
\end{subequations}
\end{satz}

\subsection{Elimination of the bath variables and the quantum Langevin 
                                                       equation} \label{lang}

The second step of the programme outlined at the beginning of this 
section is to eliminate the degrees of freedom of the heat bath assuming 
that the bath is in a thermal equilibrium state. This is achieved with
the help of the partial expectation value
\begin{align} \label{kanzust}
&\langle a_{n} \rangle_{\beta ,n}(p,q) := 
\int_{\rz^n\times\rz^n}\! \d p_{1}\d q_{1}\ldots\d p_{n}\d q_{n}\;
a_{n}(\mathbf{p},\mathbf{q}) \notag \\
&\hspace*{1.4cm}
\times \biggl(\; \prod_{j=1}^n  \frac{\omega _{j} \beta _{\mathrm{eff}} 
(\hbar\omega _{j})}{2\pi }
\;\exp\Bigl\{ - \beta _{\mathrm{eff}}(\hbar\omega _{j})\,
\bigl( p_{j}^2 /(2m_{j}) + (m_{j}\omega _{j}^2 /2) q_{j}^2 \bigr) 
\Bigr\}\biggr)
\end{align}
of an observable $ a_{n} $ of the total system. 
Consequently, we do not allow for initial correlations between the
white-noise system and the heat bath. The phase-space 
function in the second line  of (\ref{kanzust}) is the Wigner density of 
the canonical equilibrium state of the $n$ bath oscillators at inverse 
temperature $ \beta <\infty$. Note that quantum fluctuations of the heat bath
are fully taken into account through effectively increased
temperatures, defined by the function
\begin{equation}  \label{betaeff}
E \mapsto \beta _{\mathrm{eff}}(E) :=
\frac{2}{E}\; \tanh\biggl(\frac{\beta E}{2}\biggr)\,.
\end{equation}
Starting from the Heisenberg equations of motion for the $ n+1 $ 
components of position and momentum of the total system, one deduces that
$ \etln q $ satisfies a linear quantum Langevin equation
\citep{FoKaMa65,BeKa81,Maa84,FoKa87,FoLeOC88a,Gar91,JaPi97a}%
---derivatives with respect to time are symbolized by superposed dots---
\begin{equation} \label{qle}
\ddot{x}(t) + m^{-1}\ints \gamma_{n} (t-s) \,\dot{x}(s)
+ \omega ^2 x(t) + \gamma _{n}(t)\, x(0)/m = F_{n}(t)/m  
\end{equation}
with the friction kernel
\begin{equation} \label{nreikern}
\gamma _{n}(t):= \sum_{j=1}^n m_{j}\omega _{j}^2 \cos
(\omega _{j}t)
\end{equation}
and the driving force
\begin{equation} \label{nantrieb}
F_{n}(t) :=\sum_{j=1}^n  \bigl\{ m_{j}\omega _{j}^2 q_{j}
\cos (\omega _{j}t) +  \omega _{j}p_{j}\sin (\omega _{j}t)\bigr\}\,.
\end{equation}
The latter behaves with respect to the partial 
expectation value  $ \langle\winzbullet\rangle_{\beta ,n} $
as a (classical) stationary Gaussian stochastic process 
with zero mean and covariance function
\begin{equation}\label{nankov}
\langle F_{n}(t)F_{n}(t')\rangle_{\beta ,n} =
\sum_{j=1}^n \frac{m_{j}\omega _{j}^2}{\beta _{\mathrm{eff}}(\hbar\omega_{j})} 
\,\cos \bigl(\omega _{j}(t-t')\bigr)=: \varPhi_{\beta, n} (t-t') \,.
\end{equation}
Observe that the linear quantum Langevin equation (\ref{qle}) in the 
Weyl-Wigner-Moyal representation differs from the corresponding 
classical Langevin equation \citep{AnHa96,JaPi97b} only by the effective
temperatures in the covariance function of the driving force.

Thus we have shown

\begin{satz}\label{folzumikmak}
The classical phase-space trajectories generated by $ H^{(n)} $ can be
written as
\begin{subequations}\label{ges}
\begin{align}
\etln p &= m\; \partial_{t}\,\etln q\,,\label{gesp}\\
\etln q &= (p/m)\, G_{n}(t)  + q\,\dot{G}_{n}(t)
+ m^{-1} \ints G_{n}(t-s)\; F_{n}(s) \,.\label{gesq}
\end{align}
\end{subequations}
Here 
\begin{equation}\label{ngreen}
G_{n}(t):= \int_{\rz + \i\varepsilon } \frac{\d z}{2\pi }\;
\frac{\e^{-\i zt}}{\omega ^2 -z^2 - \i z\hat{\gamma }_{n}(z)/m}\,,
\qquad\qquad t\in\rz\,, \;\varepsilon  > 0\,,
\end{equation}
is the retarded Green function (or causal response function) of the quantum 
Langevin equation \eqref{qle} and 
\begin{equation}\label{stieltjes}
\hat{\gamma }_{n}(z) := \int_{0}^\infty \!\d t\; \e^{\i zt}\, 
\gamma _{n}(t) = \frac{\i}{2}\;\sum_{j=1}^n m_{j} \omega _{j}^2\;
\biggl( \frac{1}{z - \omega _{j}} + \frac{1}{z + \omega _{j}} \biggr)
\end{equation}
is the Laplace transform of the friction kernel \eqref{nreikern} defined for 
$ z\in\cz $, $ \Im z > 0 $.
\end{satz}

\begin{bems} \label{zufolzumikmak}
\item    \label{zufolzumikmakanal}
One deduces for $ \Im z > 0  $ from the explicit expression 
(\ref{stieltjes}) that
\begin{equation} 
\Im\Bigl\{ \frac{\omega ^2 - z^2}{z} \Bigr\} < 0 <
\Im \bigl\{\i \hat{\gamma }_{n}(z)\bigr\}\,,
\end{equation}
Hence, the denominator of the integrand in 
(\ref{ngreen}) has no zeros in the complex upper half-plane. 
Moreover, the integrand of (\ref{ngreen}) is 
analytic in this half-plane, and therefore the definition of the 
Green function $ G_{n} $ does not depend on $ \varepsilon $.
\item 
Note that $\lim\limits_{t\downto 0} G_{n}(t)=0$, 
$\lim\limits_{t\downto 0} \dot{G}_{n}(t)=1$
and $\lim\limits_{t\downto 0} \ddot{G}_{n}(t)=0$.
\item
The Laplace transforms of $ G_{n} $ and its derivatives  read for 
$ z\in\cz $, $ \Im z > 0 $,
\begin{subequations}
\begin{align} \label{lapgn}
\hat{G}_{n}(z) &:= \int_{0}^{\infty }\!\d t\; \e^{\i zt}\; G_{n}(t)
= \frac{1}{\omega ^2 -z^2 - \i z\hat{\gamma }_{n}(z)/m}\,, \\ 
\hat{\dot {G}}_{n}(z)  &\phantom{:}= -\i z\,\hat{G}_{n}(z)\,, 
\label{lapgnd} \\
\hat{\ddot{G}}_{n}(z)  &\phantom{:}= -z^2\,\hat{G}_{n}(z) -1\,. 
\label{lapgndd}  
\end{align}
\end{subequations}
\end{bems}

\noindent
Basically, Statement \ref{folzumikmak} reduces the analysis of the first 
summands of $ \ttnad p^2 $ and $ \ttnad q^2 $, as given by 
(\ref{raurei2}), to an analysis of the Green function $ G_{n} $.
The next statement does the same for the second summands in
(\ref{raurei2}).

\begin{satz}\label{omgam}
Let $ \mathcal{S}_{\delta}:= \{ z\in\cz : |\Im z| <\delta \} $ 
be a strip of width $ 2\delta > 0 $ centred around the real axis in the 
complex plane and let $ f: \mathcal{S}_{\delta}\longrightarrow \cz $ be 
a bounded analytic function on $ \mathcal{S}_{\delta} $ with reflection
symmetry $ f(z)=f(-z) $ for all $ z\in\mathcal{S}_{\delta} $. 
Then, for any $0 < \varepsilon < \delta $, 
the $ 00 $-matrix element $\bigl( f(\Omega )\bigr)_{00}$, where
$ \Omega $ stands for the positive square root of 
the $ (n+1)\times(n+1) $-matrix of squared frequencies \eqref{ommat}, 
may be expressed in terms of a principal-value integral
\begin{equation}\label{omgamglg}
\bigl( f(\Omega )\bigr)_{00} = \frac{1}{\pi\i}\;
\valprinc\int_{\rz + \i\varepsilon }\!\d z\;
z\,f(z)\,\hat{G}_{n}(z)\,.
\end{equation}
\end{satz}

\begin{proof}
We closely follow \citet{Ull66} and \citet{FoLeOC88b}.
Using the spectral representation
$ \Omega = \mathsf{U}\,\mathrm{diag}(\Omega _{0},\ldots,\Omega _{n})
\mathsf{U}^{\mathsf{t}}  $,
where $ \mathsf{U}$ is an orthogonal $ (n+1)\times (n+1) $-Matrix  
and $ 0\le\Omega _{0}\le\Omega _{1}\le\ldots\le\Omega _{n} $ 
are the eigenvalues of $ \Omega $, the left-hand side of 
(\ref{omgamglg}) is rewritten as
\begin{equation} \label{spekreihe}
\bigl( f(\Omega )\bigr)_{00} = \sum_{k=0}^n (\mathsf{U}_{0k})^2\, 
f(\Omega _{k})\,.
\end{equation}
The proof of the following claim is contained in \citet{FoLeOC88b}.

\vspace{2ex}\noindent
\itshape The function
\begin{equation} \label{alphadef}
\xi \mapsto\alpha (\xi ):= \hat{G}_{n}(\sqrt{\xi }) =
\frac{1}{\omega ^2 - \xi  - \i\sqrt{\xi }\,
\hat{\gamma }_{n}(\sqrt{\xi })/m}
\end{equation}
extends to an analytic function on 
$ \cz \setminus \{ \Omega _{0}^2,\ldots,\Omega _{n}^2 \} $. 
Moreover, $ \Omega _{j} \neq \Omega _{k} $ for $ j\neq k $, and 
$ \alpha $ has a simple pole at $ \xi =\Omega _{k}^2 $ with residue
$ \mathrm{Res}_{\Omega _{k}^2}\{\alpha \} = - (\mathsf{U}_{0k})^2 $
for $ k=0,\ldots,n $. \upshape

\vspace{3ex}\noindent
Due to the symmetry $ f(z)=f(-z) $ there exists an analytic function $ g $
on $ \mathcal{S}_{\delta}^2 :=  \{ \xi \in\cz : \xi =z^2, 
z\in\mathcal{S}_{\delta}\}  $ such that $ g(z^2)=f(z) $.
Hence, the above claim and (\ref{spekreihe}) lead in combination with 
the residue theorem to the representation
\begin{figure}%
\unitlength.8cm
\begin{picture}(13,19)
\put(-0.3,10.3){                
    \begin{picture}(13,8)         
    \put(12.2,3.5){\small Re $ \xi  $}
    \put(4.7,7.2){\small Im $ \xi  $}
    \put(1,4){\vector(1,0){12}}
    \put(4.5,0.5){\vector(0,1){7}}
    \put(1,7){(a)}
    \put(4.37,4){
        \begin{picture}(8.5,4)
        \thicklines
        \put(0.5,0){\circle*{0.13}}
        \put(1.1,0){\circle*{0.13}}
        \put(1.9,0){\circle*{0.13}}
        \put(3.1,0){\circle*{0.13}}
        \put(4.5,0){\circle*{0.13}}
        \put(0.35,-0.5){\small $\Omega _{0}^2$}
        \put(2.1,-0.5){$ \cdot~\cdot~\cdot $}         
        \put(4.35,-0.5){\small $\Omega _{n}^2$}
  
        \put(2.5,0){\oval(6,1.4)}
        \put(2.5,0.7){\vector(1,0){0.1}}
        \put(2.5,-0.7){\vector(-1,0){0.1}}
        \put(3.5,0.8){\small $ {K} $}
        \put(-1.2,-0.5){\small $-\varepsilon ^2$}
        \put(5.7,-0.5){\small $\ell^2$}

        \thinlines
        \qbezier[70](-1.5,0)(-1.5,2.5)(7.5,3)
        \qbezier[70](-1.5,0)(-1.5,-2.5)(7.5,-3)
        \put(7,2.3){\small $ \mathcal{S}_{\delta}^2 $}
        \put(-2.2,-0.5){\small $-\delta ^2$}
        \end{picture}
        }
    \end{picture}
    }
\put(-0.3,.8){                     
    \begin{picture}(13,8)
    \put(12.2,3.5){\small Re $ z  $}
    \put(6.75,7.2){\small Im $ z  $}
    \put(1,4){\vector(1,0){12}}
    \put(6.55,0.5){\vector(0,1){7}}
    \put(1,7){(b)}
    \put(6.4,4){
        \begin{picture}(8.5,4)
        \thicklines
        \multiput(0.6,0)(.7,0){5}{\circle*{0.13}}
        \put(0.45,-0.5){\small $\Omega _{0}$}
        \put(1.6,-0.5){$ \cdot~\cdot~\cdot $}         
        \put(3.25,-0.5){\small $\Omega _{n}$}

        \multiput(-0.6,0)(-.7,0){5}{\circle*{0.13}}
        \put(-1.05,-0.5){\small $-\Omega _{0}$}
        \put(-2.4,-0.5){$ \cdot~\cdot~\cdot $}         
        \put(-3.87,-0.5){\small $-\Omega _{n}$}
  
        \put(-4.3,0.8){\line(1,0){8.7}}
        \put(1.9,0.8){\vector(1,0){0.1}}
        \put(-1.9,0.8){\vector(1,0){0.1}}
        \put(0.8,0.95){\small $ K_{2} $}
        \put(-0.28,0.95){\small $ \varepsilon  $}
  
        \put(-4.3,0){\line(0,1){0.8}}
        \put(-4.3,0.4){\vector(0,1){0.1}}
        \put(-5,0.3){\small $ K_{1} $}
        \put(-4.63,-0.5){\small $ -\ell $}
  
        \put(4.4,0){\line(0,1){0.8}}
        \put(4.4,0.4){\vector(0,-1){0.1}}
        \put(4.6,0.3){\small $ K_{3} $}
        \put(4.34,-0.5){\small $ \ell $}

        \thinlines
        \qbezier[70](-5,2)(0,2)(5.5,2)
        \qbezier[70](-5,-2)(0,-2)(5.5,-2)
        \put(5,1.5){\small $ \mathcal{S}_{\delta} $}
        \put(-0.28,2.1){\small $ \delta $}
        \end{picture}
        }
    \end{picture}
    }
\end{picture}
\vspace*{.2cm}
\caption{%
The poles of the integrand and the integration contour for  
(a) the integral (\protect{\ref{xiint}}) and
(b) the integral (\protect{\ref{zint}}).
}%

%
\label{ebene}%
\end{figure}
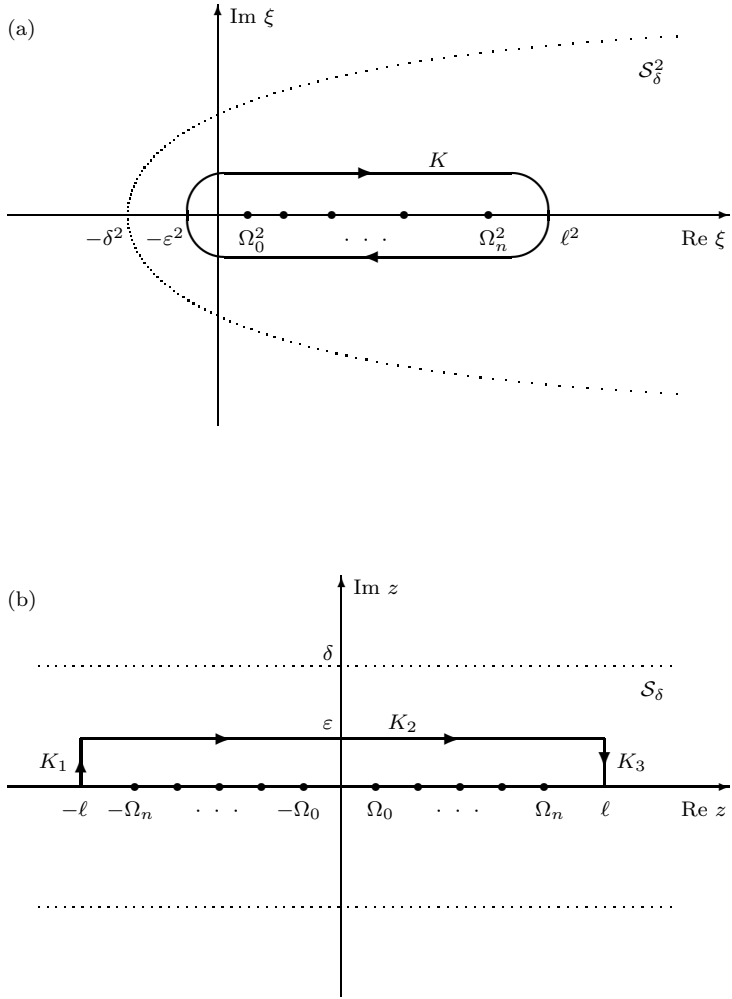%
\begin{equation} \label{xiint}
\bigl( f(\Omega )\bigr)_{00} = \frac{1}{2\pi \i}\int_{K}\!
\d\xi \; g(\xi )\,\alpha (\xi )\,.
\end{equation}
As is depicted in Figure \mbox{\ref{ebene}(a)}, the integration contour
$ K $ encloses clockwise all poles of $ \alpha $.
It intersects the real axis at 
$ \xi =-\varepsilon ^2 $ and at $ \xi =\ell^2 $, where
$ 0<\varepsilon <\delta $  and $ \ell > \Omega _{n} $.
Next we introduce $ z:=\sqrt{\xi } $ as the square root of 
$ \xi $ with branch cut along the positive real half-axis 
and perform a change-of-variables in the integral (\ref{xiint}) 
\begin{equation} \label{zint}
\bigl( f(\Omega )\bigr)_{00} = \frac{1}{\pi\i}\int_{K_{1}+K_{2}+K_{3}}\!
\d z \; z\,f(z)\,\alpha (z^2 )\,,
\end{equation}
see Figure \mbox{\ref{ebene}(b)}. In the limit $ \ell\to\infty $
the contributions of the vertical line segments $ K_{1} $ and 
$ K_{3} $ to the integral (\ref{zint}) vanish due to the boundedness of 
$ f $, whereas the contribution of the horizontal line segment 
$ K_{2} $ converges to the principal-value integral in
(\ref{omgamglg}).
\end{proof}

\subsection{The macroscopic limit and the reduced averaged dynamics}
                                                              \label{mac} 

In order to enable the white-noise system to dissipate energy into the
heat bath, it is necessary to take the macroscopic limit of the heat
bath such that a continuum of frequencies arises.
More precisely, guided by the relation 
\begin{equation}
  \gamma_{n}(t) = \int_{\rz}\!\d\nu\; J_{n}(\nu)\,\e^{\i\nu t}\,,
\end{equation}
where $ J_{n}(\nu):= \frac{1}{2}\,\sum\limits_{j=1}^{n} m_{j}\omega_{j}^{2}
\bigl[ \delta(\nu - \omega_{j}) + \delta(\nu + \omega_{j})\bigr]$,
we require that with the number $ n $ of bath 
oscillators going to infinity, the limit of the friction kernel 
$ \gamma (t) := \lim\limits_{n\to\infty } \gamma _{n}(t)$
exists and admits a representation as the Fourier transform of the so-called
spectral density $ J $ of the macroscopic heat bath
\begin{equation}\label{dos}
\gamma (t) = \int_{\rz}\!\d\nu \; J(\nu )\,\e^{ \i\nu t}\,.
\end{equation}
The spectral density describes the strength of the coupling between
the system and the heat bath. In what follows, it is supposed to satisfy

\begin{annahmen} \label{jan}
\item   \label{jannull}
$J$ is continuous, even and strictly positive, ~$ J(\nu ) = J(-\nu ) >
0 $ ~for all $ \nu \in\rz\, $.
\item   \label{jananal}
There are two constants $ 0 < \psi < \pi /2 $ and $ \delta > 0 $
such that $ J $ admits an analytic continuation to the subset
\begin{figure}%
\unitlength.8cm
\begin{picture}(13,5.5)
\put(12,3.75){\small Re $ z  $}
\put(7.05,4.7){\small Im $ z  $}
\thicklines
\put(.8,3.5){\vector(1,0){12.1}}
\thinlines
\put(6.85,0.5){\vector(0,1){4.5}}
\put(6.75,3.5){
    \begin{picture}(8.5,4)
    \put(-4.2,-1.1){\small $ \mathcal{H}^{-}_{\psi ,\delta }  $}
    \put(-0.8,-.7){\small $ -\delta $}
    \qbezier[26](0,-.8)(1.3,-.8)(2.6,-.8)               
    \qbezier[8](2.15,-1.5)(2.45,-1.15)(2.3,-.8)         
    \put(1.5,-1.18){\small $ -\psi $}
    \renewcommand{\xscale}{-.61}
    \renewcommand{\yscale}{-.61}
    \renewcommand{\xscaley}{.81}
    \renewcommand{\yscalex}{-.81}
    \put(2.15,-1.5){\scaleput(0,0){\curve[5](0,0, -.18,-.08)}}  
    \put(2.15,-1.5){\scaleput(0,0){\curve[5](0,0, -.18,.08)}}   
    \thicklines
    \put(0,-.8){\line(3,-1){6}}
    \put(0,-.8){\line(-3,-1){6}}
    \end{picture}
}
\end{picture}
\caption{%
The subset \protect{$ \mathcal{H}^{-}_{\psi ,\delta } := \bigl\{ z\in \cz : 
\Im z < 0,\, \arg (z+\i\delta ) \protect\protect\protect\notin 
[  - \pi + \psi , - \psi ] \, \bigr\} $} of the complex plane.}%

%
\label{Kform}%
\end{figure}
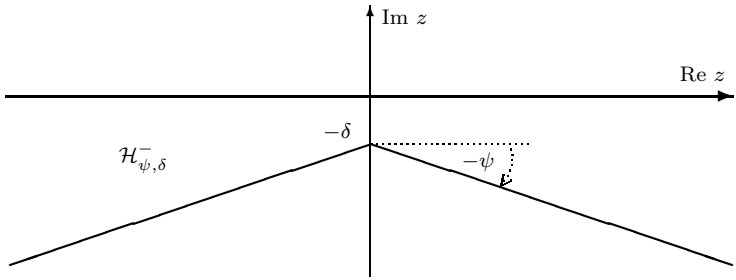%
$ \mathcal{H}^{-}_{\psi ,\delta } := \bigl\{ z\in\cz : \Im z < 0 ~\mathrm{and}~
\arg (z+\i\delta ) \notin [ -\pi +\psi , -\psi ] \,\bigr\} $ 
of the complex lower half-plane, see Figure \ref{Kform}.
\item   \label{janweg}
$J$ has the following decay properties:\\
$ \sup\limits_{z\in\mathcal{H}^{-}_{\psi ,\delta }}\bigl| z^{2}J(z)\bigr|
<\infty\,$ and the limit 
$ L:= \lim\limits_{|\nu |\to\infty } \nu ^2 \, J(\nu ) $, taken along 
the real axis, exists and is approached sufficiently  fast,
\begin{equation} \label{abfallreell}
\lim_{|\nu |\to\infty } \bigl\{ |\nu |^r \bigl( \nu^2\, J(\nu ) - L \bigr)
\bigr\} = 0 \qquad\quad \text{for~some~~} r > 0\,.
\end{equation}
\end{annahmen}

\noindent
Note that Assumptions \ref{jannull} and \ref{janweg} imply the
integrability of $J$ over $\rz$.

Assumptions \ref{jan} are obviously satisfied for the Gaussian spectral density
$ J(\nu )= J_{0} \exp\bigl\{ -\nu ^2/(2\omega _{0}^2)\bigr\} $,
$ J_{0},\,\omega _{0} > 0 $. Another admissible choice for $ J $ is the
Cauchy-Lorentz spectral density $ J(\nu )= J_{0}\,\omega _{0}^{2} / 
( \omega _{0}^2 + \nu ^2 ) $ of the so-called Drude model. 
Both densities approximate Ohmic damping, which is defined by
$J(\nu)=J_{0}$ for all $\nu$, the better the larger $\omega_{0}$ is.
Ohmic damping itself is excluded by the decay assumption 
\ref{janweg}. It leads for $\hbar \neq 0$ to some unphysical
ultraviolet divergencies, see for example \eqref{ankov} and \eqref{theequp}
below. The reader should be aware that the recent
attempt of \citet{FoOC96} to interpret some of these divergencies in a
na{\"\i}ve distributional sense has failed: the proposed distributional 
derivative of the hyperbolic cotangent makes only sense for test functions 
which vanish at the origin. But this implies that the distributional 
derivative does not differ from the classical one.

For later purpose we state that the covariance function
(\ref{nankov}) of the driving force $ F_{n} $ in the quantum 
Langevin equation evaluates in the macroscopic limit to
\begin{equation}\label{ankov}
\varPhi_{\beta} (t):= \lim_{n\to\infty } \varPhi_{\beta, n} (t)
= (\hbar/2) \int_{\rz}\!\d\nu \; J(\nu )\,\nu \,
\coth(\beta \hbar\nu /2)\,\e^{\i\nu t}\,.
\end{equation}
Accordingly, its Fourier transform  
differs from that of the friction kernel simply by the factor of
$1/\beta_{\mathrm{eff}}(\hbar\nu)$, a fact which reflects the
fluctuation-dissipation theorem \citep{KuToHa91}. 
Furthermore we use the notation 
\begin{equation} \label{green}
G(t):= \lim_{n\to\infty } G_{n}(t) 
\end{equation}
for the macroscopic limit of the Green function of the quantum 
Langevin equation. It does not only depend on the spectral density $J$
of the heat bath, but also on the mass $m$ and the frequency $\omega$ 
of the system Hamiltonian.

After these preparations and recalling the partial expectation value
(\ref{kanzust}) with respect to the canonical equilibrium state at 
inverse temperature $ \beta $ of the heat bath, the reduced averaged 
dynamics of system observables on phase space $ \rxr $ is defined as
\begin{equation} \label{reddef}
\rtad := \lim_{n\to\infty }
\bigl\langle \ttnad \winzbullet\bigr\rangle_{\beta ,n} \,,
\end{equation}
confer \eqref{gemgesdyn} and \eqref{kanzust}. Its adjoint $ \rt $ with 
respect to the standard scalar product for 
square-integrable functions on phase space $ \rxr $ describes the 
reduced averaged dynamics for the states of the white-noise system. 

The starting point for deriving the long-time spreading of states
under this dynamics is

\begin{satz}\label{reddyn}
\begin{subequations}\label{red2}
\begin{gather}  \label{redp2} 
\begin{split} 
\rtad p^2 = \bigl(p\,\dot{G}(t) + mq\,\ddot{G}(t)\bigr)^2
&+ \ints\!\intsp \dot{G}(s)\,\varPhi_{\beta} (s-s')\,\dot{G}(s')\\
& + (-\partial_{q}^2 C)(0)\ints \bigl(\dot{G}(s)\bigr)^2 \,,
\end{split}  \\   \label{redq2} 
\begin{split} 
\rtad q^2 = \Bigl(\frac{p}{m}\, G(t) + q\,\dot{G}(t)\Bigr)^2
&+ m^{-2} \ints\!\intsp G(s)\,\varPhi_{\beta} (s-s') \,G(s')\\
&+ m^{-2}(-\partial_{q}^2 C)(0)\ints \bigl( G(s)\bigr)^2\,.
\end{split}
\end{gather}
\end{subequations}
\end{satz}

\begin{proof}
The terms in the first line of (\ref{redp2}) and (\ref{redq2}), 
respectively, follow 
from taking the partial expectation value (\ref{kanzust}) of the first 
term in (\ref{raureip}) and (\ref{raureiq}), respectively. It is
evaluated in the macroscopic limit by using
Statement \ref{folzumikmak}, as well as (\ref{nankov}), (\ref{ankov})
and (\ref{green}).
The term in the second line of (\ref{redq2}) emerges in the macroscopic 
limit from the second term in (\ref{raureiq}) due to 
\begin{equation} \label{zwischenint}
\biggl( \frac{\sin(\Omega s)}{\Omega } \biggr)_{00}
= - \int_{\rz + \i\varepsilon }\frac{\d z}{2\pi }\;
\bigl(\e^{\i zs} - \e^{-\i zs}\bigr)\, \hat{G}_{n}(z)
= G_{n}(s)\,, \qquad s \ge 0\,.
\end{equation}
Here we have applied Statement \ref{omgam} and used the causality 
condition
\begin{equation}   \label{cauchyres}
G_{n}(s) =0 \quad \text{for all}\quad s\le0\,,
\end{equation}
which is a consequence of Cauchy's integral theorem for analytic 
functions and Definition (\ref{ngreen}). As to the term in the 
second line of (\ref{redp2}), we simply refer to (\ref{zwischenint})
and to $ \cos(\Omega s) = \partial_{s} \sin(\Omega s)/\Omega $.
\end{proof}

\subsection{The long-time limit} \label{longtime}

The dissipative character of the reduced averaged dynamics will become 
obvious from the long-time behaviour of the Green function $ G $ 
and of the covariance function $ \varPhi_{\beta} $. In fact, under our 
Assumptions \ref{jan} these functions exhibit an exponential relaxation.

\begin{satz} \label{greentotik}
There is a decay constant $\eta >0$ such that the
long-time behaviour of the macroscopic Green function \eqref{green}
and of its derivatives is given by
\begin{align} 
G(t) &=  
\Biggl\{\begin{array}{@{\;}l@{\quad}l}    
\bigl( \pi J(0)\bigr)^{-1} + 
\mathcal{O}(\e^{-\eta t})\,, & \text{if} \quad \omega = 0\,,\\[.7ex]
\mathcal{O}(\e^{-\eta t})\,, & \text{if} \quad \omega > 0\,, 
\end{array} \label{as}\\
\dot{G}(t) &= \phantom{\biggl\{\;}
\mathcal{O}(\e^{-\eta t}) \,,  \label{dotas}\\
\ddot{G}(t) &=\phantom{\biggl\{\;}
\mathcal{O}(\e^{-\eta t}) \,. \label{ddotas}
\end{align}
The covariance function $ \varPhi_{\beta} $ from \eqref{ankov} is Lebesgue
integrable over $ \rz $. It decays exponentially for $ |t| \to\infty $.
For $ t \to 0 $ it either remains finite or diverges logarithmically.
\end{satz}

\begin{proof}
  The Assumptions \ref{jan} on the spectral density $J$ translate into
  analyticity and other properties of $\hat{\gamma}$. We summarize them in an 
  \par \vspace{1ex} \noindent
  \itshape Assertion.\quad There are two constants $0<\psi<\pi/2$ and
  $0<\eta<\delta$ such that $\hat{\gamma}$ can be continued
  analytically from the upper complex half-plane to the sector
  $\mathcal{H}_{\psi,\eta}:= \mathcal{H}^{-}_{\psi,\eta}
  \cup \{z\in\cz : \Im z \ge 0\}$. Moreover,
  \begin{equation} \label{abfallfort}
    \sup_{z \in \mathcal{H}_{\psi ,\eta }} | z\,\hat{\gamma }(z) |
    < \infty \,,
  \end{equation}
  and the complex function
  \begin{equation} \label{nullstellen}
    z \mapsto \omega ^2 -z^2 -\i z\,\hat{\gamma }(z)/m 
  \end{equation}
  has no zeros in $\mathcal{H}_{\psi,\eta}$ ---except for
  the case $\omega =0$ in which $z=0$ is a simple zero.
  \par\vspace{2ex plus .5ex minus .5ex} \upshape\noindent
  Before we show the validity of the Assertion, we demonstrate how it
  can be exploited to prove the Statement. Consider the inverse
  Laplace transformation of the functions $ \hat{\dot{G}}(z) = -\i
  z\hat{G}(z)  $, $ \hat{\ddot{G}}(z) = -z^2\hat{G}(z) -1 $ and
  $\hat{G}(z)$ (in the case $\omega > 0$), respectively 
  $ \hat{G}(z) + \bigl( \i z\,\hat{\gamma }(0)/m \bigr)^{-1} $  (in the
  case $\omega =0$). For the last function to make sense, observe that
  $ \hat{\gamma }(0)= \pi J(0) > 0 $ due to Assumption
  \ref{jannull}. According to the Assertion each of these functions is
  analytic in $\mathcal{H}_{\psi,\eta}$, remains bounded for 
  $ z \to -\i\eta $ and decays like $ \mathcal{O}(|z^{-1}|) $
  for $ |z|\to\infty $, $ z\in\mathcal{H}_{\psi ,\eta }$.  
  Therefore Cauchy's theorem allows to deform the integration
  contour in the inverse Laplace transformation from the upper
  complex half-plane to the boundary $\partial\mathcal{H}_{\psi
  ,\eta }$. But now we are in the situation of the Abelian
  theorem in \citet[Chap.\ 15, \S 4, Thm.~2]{Doe50}, and 
  \eqref{as}--\eqref{ddotas} follow. 

  To prove the above assertion observe that 
  \begin{equation} \label{stieltjesfort}
    \hat{\gamma }(z) := \i \int_{\partial\mathcal{H}_{\psi ,\delta}}\!\d z'\;
    \frac{J(z')}{z-z'}\,,\qquad\quad z\in \mathcal{H}_{\psi ,\eta }\,,
    \; 0 < \eta < \delta \,,
  \end{equation}
  coincides in the upper half-plane $\Im z>0$ with the former
  definition (see \eqref{stieltjes} in the macroscopic limit)
  \begin{equation}  \label{laplacetrafo}
    \hat{\gamma }(z) = \int_{0}^\infty \!\d t\; \e^{\i zt}\, \gamma (t)
    = \i \int_{\rz}\!\d\nu \; \frac{J(\nu )}{z-\nu }\,.
  \end{equation}
  This is due to Assumptions \ref{jananal} and \itemref{janweg} and
  Cauchy's theorem. Moreover, for $z_{0}\in\mathcal{H}_{\psi,\eta}$
  the difference quotient $\bigl(\hat{\gamma }(z) - \hat{\gamma}
  (z_{0})\bigr)/(z-z_{0})$ exists in the limit $ z\to z_{0} $ by Assumption
  \ref{janweg} and dominated convergence. Hence, \eqref{stieltjesfort}
  provides the desired analytic continuation of $\hat{\gamma}$.
  The decay property \eqref{abfallfort} is evident from
  \eqref{stieltjesfort} and Assumption \ref{janweg}. A necessary
  condition for a zero of \eqref{nullstellen}---except for the
  zero $z=0$ in the case $\omega =0$ ---is $\Im z < 0$ as follows from
  Remark \ref{zufolzumikmakanal},  $\Re\hat{\gamma}(\nu) = \pi J(\nu)>0$
  for $\nu\in\rz$ and Assumption \ref{jannull}. Furthermore, \eqref{abfallfort}
  implies that all zeros lie in a bounded region of
  $\mathcal{H}^{-}_{\psi,\eta}$. Hence, there exist only
  finitely many of them, because otherwise they would have a limit point, and
  by the identity theorem for analytic functions the function
  \eqref{nullstellen} would have to vanish for all
  $z\in\mathcal{H}_{\psi,\eta}$. One can
  therefore choose $\psi$ and $\eta$ sufficiently small such
  that there are no zeros at all in
  $\mathcal{H}_{\psi,\eta}$. This completes the proof of the
  Assertion.

  It remains to prove the properties of the covariance function 
  $\varPhi_{\beta}$. Because of 
  \begin{equation} \label{gammaint}
    \varPhi_{\beta} (-t) = \varPhi_{\beta} (t) 
    = \hbar \,\Re\biggl\{ \int_{0}^\infty\!\d\nu\; J(\nu ) 
    \,\nu\, \coth (\beta \hbar\nu /2)\; \e^{-\i\nu t}\biggr\} 
  \end{equation}
  we assume $t\ge 0$ from now on. As a consequence of
  \eqref{abfallreell}, $J(\nu)$ either decays like $\nu^{-2}$ for 
  $\nu\to\infty$ or it is of the order $\mathcal{O}(\nu^{-2-r})$ for some
  $r>0$. In the latter case the integral \eqref{gammaint} exists as a
  Lebesgue integral for all $t\ge 0$ rendering $\varPhi_{\beta}$ a
  bounded function. In the former case the integral \eqref{gammaint}
  exists as an improper Riemann integral for $t>0$, and
  $\varPhi_{\beta}$ has a singularity at $t=0$. In order to
  characterize this singularity, fix some $s>0$, set
  $L:=\lim\limits_{\nu\to\infty} \bigl(\nu^{2}\,J(\nu)\bigr)$ and make the
  decomposition 
  \begin{equation}                    
    J(\nu )\,\nu\,\coth (\beta \hbar\nu /2) =: \frac{L}{\nu } - 
    \frac{L}{\nu }\;\e^{-s\nu }  + f(\nu )
  \end{equation}
  which defines the Lebesgue integrable function $f$ over
  $[0,\infty[$. Hence, $f$ does not contribute to the singularity of 
  $\varPhi_{\beta}$ at $t=0$. The singularity of $\varPhi_{\beta}$
  comes solely from 
  \begin{equation} 
    \int_{0}^\infty \!\d\nu\;\biggl( \frac{L}{\nu } -
    \frac{L}{\nu }\;\e^{-s\nu } \biggr)\; \e^{-\i\nu t}\,,
  \end{equation}
  and the Abelian theorem in \citet[Chap.\ 13, \S 3, Thm.~1]{Doe50}
  reveals that it is logarithmic.
  
  In order to derive the exponential decay of $\varPhi_{\beta}$
  for $t\to\infty$ we proceed as in the case of the Green
  function. Analyticity and decay of $J$ in
  $\mathcal{H}^{-}_{\psi,\eta}$ imply together with Cauchy's
  theorem the representation
  \begin{align} 
    \varPhi_{\beta} (t) &= (\hbar/2) \int_{\rz}\!\d\nu \; J(\nu )\,\nu\, 
    \coth (\beta \hbar\nu /2)\;\e^{-\i\nu t}\notag\\
    &= (\hbar/2) \int_{\partial\mathcal{H}_{\psi ,\eta }}\!\d z\; 
    J(z)\, z\, \coth (\beta \hbar z /2)\;\e^{-\i zt}\,,
  \end{align}
  which is valid for $t>0$ and $ 0 < \eta < \min \bigl\{ \delta
  , 2\pi /(\beta \hbar) \bigr\} $. The Abelian theorem
  \citep[Chap.\ 15, \S 4, Thm.~2]{Doe50} then yields the exponential
  decay of $\varPhi_{\beta}$ for $t\to\infty$. Thus
  $\varPhi_{\beta}$ is also seen to be Lebesgue integrable over $\rz$.
\end{proof}

\noindent
To formulate the main result of this section, it is
illuminating to study first the expectation value 
\begin{equation} \label{gleichges}
\gesgler{a} := \lim_{n\to\infty } \langle w^{(n)} , a \rangle
\end{equation}
of a system observable $a$ with respect to the thermal equilibrium
state of the deterministic part $H^{(n)}$ of the \emph{total} Hamiltonian
in the macroscopic limit of the heat bath. The scalar product in
\eqref{gleichges} was defined in \eqref{nskalar}. Recalling the
effective temperatures \eqref{betaeff}, the Wigner 
density of the corresponding pre-limit state may be written as 
\begin{multline} 
w^{(n)}(\mathbf{p},\mathbf{q}):= 
\frac{\det\bigl( \Omega \,
\beta_{\mathrm{eff}}(\hbar\Omega ) \bigr)}{(2\pi )^{n+1}}\;
\exp\Bigl\{ - \tfrac{1}{2} \; \mathbf{p}\cdot\mathsf{M}^{-1/2}
\beta_{\mathrm{eff}}(\hbar\Omega )\mathsf{M}^{-1/2}\mathbf{p} \\
- \tfrac{1}{2}\; \mathbf{q}\cdot\mathsf{M}^{1/2}\Omega ^2 
\beta_{\mathrm{eff}}(\hbar\Omega )\mathsf{M}^{1/2}\mathbf{q}
\Bigr\}\,.
\end{multline}
Statement \ref{theequ} below compiles three equivalent expressions for the
thermal expectation values $\gesgler{p^{2}}$ and $\gesgler{q^{2}}$, which are
available in the literature, see for example 
\citet{HaRe85}, \citet{FoLeOC88b}, \citet[Table~2]{GrScIn88}
and \citet[Chap.~6]{Wei93}. Nevertheless we will prove them, since Statement
\ref{omgam} provides a quick way to do so.

\begin{satz}  \label{theequ}
The following identities hold under the Assumptions \textup{\ref{jan}} 
on the spectral density:
\begin{subequations} \label{theequpq}
\begin{align} \label{theequp}
\gesgler{p^2} &= \frac{\hbar}{2}\; \int_{\rz}\!\d\nu\; J(\nu )\, 
\nu^3 \, \bigl| \hat{G}(\nu )\bigr|^2 \, 
\coth (\beta \hbar\nu /2)\notag \\
&= \frac{m \hbar}{2\pi \i}\; \valprinc\int_{\rz}\!\d\nu\;  
\bigl( \nu ^2 \hat{G}(\nu ) +1\bigr) \,  
\coth (\beta \hbar\nu /2)\notag \\
&= \frac{m}{\beta }\; \sum_{l=-\infty }^{+\infty } \bigl( 1- \nu _{l}^2
\hat{G}(\i |\nu _{l}|)\bigr)
\end{align}
and, if $\omega > 0$,
\begin{align} \label{theequq}           
\gesgler{q^2} &= \frac{\hbar}{2m^2}\; \int_{\rz}\!\d\nu\; J(\nu )\,
\nu\,\bigl| \hat{G}(\nu )\bigr|^2 \,
\coth (\beta \hbar\nu /2)\notag\\
&= \frac{\hbar}{2\pi \i m}\; \valprinc\int_{\rz}\!\d\nu \;
\hat{G}(\nu )\,\coth (\beta \hbar\nu /2)\notag\\
&= \frac{1}{m\beta}\; \sum_{l=-\infty }^{+\infty } 
\hat{G}(\i |\nu _{l}|)\,.
\end{align}
\end{subequations}
Here, $\nu _{l}:= 2\pi l/(\beta \hbar)$ are the even Matsubara 
frequencies, $\hat{G}(\nu )$ denotes the analytic continuation of 
$\hat{G}(z)$ to the real axis, and the principal-value prescription in
the middle lines of \eqref{theequp} and \eqref{theequq} has to be 
applied at the origin.
\end{satz}

\begin{bem}  \label{zutheequ}
  As is indicated by the notation, the expectation values
  $\gesgler{p^2}$ and $\gesgler{q^2}$ depend on the inverse
  temperature $\beta$ and on the spectral density $J$ of the heat
  bath. Moreover, they do also depend on Planck's constant $\hbar$.  
  From the series representations in \eqref{theequpq} one can derive  
  the following chain of inequalities
  \begin{align}
    \max\Bigl\{ \frac{1}{2\beta}, \frac{(\omega/\omega_J)^2}%
    {2\beta_{\mathrm{eff}}(\hbar\omega_J)}\Bigr\}
    &\le \frac{m\omega^2}{2}\,\gesgler{q^2}
    \le \frac{1}{2\beta_{\mathrm{eff}}(\hbar\omega)} \notag\\ 
    &\le \frac{1}{2m}\,\gesgler{p^2} 
    \le \frac{1}{2\beta_{\mathrm{eff}}(\hbar\omega_J)}\,,
  \end{align}
  where $\omega_J := \sqrt{\omega^2 + \gamma (0)/m}$ and 
  $\gamma (0) = \int_{\rz}\!\d\nu\; J(\nu ) $.
  As a consequence, one can easily deduce the values of $\gesgler{p^2}/2m$
  and $m\omega^2 \gesgler{q^2}/2$ in the low-friction limit $J\downto 0$
  or the classical limit $\hbar\downto 0$. For $J\downto 0$ both expectation
  values tend to the quantum-mechanical canonical equilibrium value 
  $1/2\beta_{\mathrm{eff}}(\hbar\omega)$ with respect to the 
  harmonic-oscillator Hamiltonian $ p^2/(2m) + (m\omega ^2/2)q^2 $. 
  For $\hbar\downto 0$ they tend to $1/2\beta$ for all admissible $J$,
  as dictated by the equipartition theorem.
  The latter expression emerges also in the quantum-mechanical case 
  $ \hbar\neq 0 $ as an asymptotic value for high temperatures 
  $\beta \downto 0$.
\end{bem}

\begin{proof}[Proof\phantom{x}of\phantom{x}Statement \ref{theequ}]
In order to calculate $\gesgler{p^2}$ one has to consider the 
macroscopic limit of the expectation value
\begin{equation} \label{pnmat}
\langle w^{(n)}, p^2 \rangle = \frac{m\hbar}{2}\,
\bigl(\Omega \coth(\beta \hbar\Omega /2) \bigr)_{00}\,.
\end{equation}
Before doing this we rewrite the right-hand side of \eqref{pnmat} with 
the help of Statement \ref{omgam}. To this end, observe that the 
function
\begin{equation} 
z \mapsto f(z)= \frac{z}{1+ \lambda ^{-2} z^2} \;\coth (\beta \hbar z/2)
\end{equation}
fulfills the assumptions of Statement \ref{omgam} for all 
$ 0 < \varepsilon < 2\pi /(\beta \hbar) < \lambda $. Hence, 
\begin{align} 
\langle w^{(n)}, p^2 \rangle &=  \frac{m\hbar}{2\pi\i}\;
\lim_{\lambda \to\infty }\;\int_{\rz + \i\varepsilon }\!\d z\;
\frac{1}{1+ \lambda ^{-2} z^2}\; z^2 \hat{G}_{n}(z) \,
\coth (\beta \hbar z/2 ) \notag\\
&= \frac{m\hbar}{2\pi\i}\; \int_{\rz + \i\varepsilon }\!\d z\;
\bigl( z^2 \hat{G}_{n}(z)  +1 \bigr)\,\coth (\beta \hbar z/2 ) \notag\\
&\phantom{=}\; -  \frac{m\hbar}{2\pi\i}\;
\lim_{\lambda \to\infty }\;\int_{\rz + \i\varepsilon }\!\d z\;
\frac{\coth (\beta \hbar z/2 )}{1+ \lambda ^{-2} z^2}\,,
\end{align}
where we have taken advantage of the fact that $z^2 \hat{G}_{n}(z)+1$
falls of like $|z|^{-2}$ for $|z|\to\infty$. Thanks to 
\begin{align} 
\int_{\rz + \i\varepsilon }\!\d z\;
\frac{\coth (\beta \hbar z/2 )}{1+ \lambda ^{-2} z^2} 
&= \;\valprinc\int_{\rz}\!\d\nu \;
\frac{\coth (\beta \hbar \nu /2 )}{1+ \lambda ^{-2} \nu^2} \;
- \pi\i  \Res_{z=0}\bigl\{ \coth (\beta \hbar z/2 ) \bigr\}\notag\\
&= - 2 \pi \i /(\beta \hbar) 
\end{align}
we conclude 
\begin{align} \label{ppre}
\gesgler{p^2} &= \frac{m\hbar}{2\pi\i} \biggl\{ 
\int_{\rz + \i\varepsilon }\!\d z\; 
\bigl( z^2 \hat{G}(z)  +1 \bigr)\,\coth (\beta \hbar z/2 ) 
\; + 2 \pi \i /(\beta \hbar) \biggr\} \notag\\
&= \frac{m\hbar}{2\pi\i} \;\valprinc\int_{\rz}\!\d \nu\; 
\bigl( \nu^2 \hat{G}(\nu )  +1 \bigr)\,\coth (\beta \hbar \nu /2 )\,.
\end{align}
The principal-value prescription has to be taken at the origin, 
and $\hat{G}(\nu )$ denotes the analytic continuation of $\hat{G}(z)$ to 
the real axis. This continuation exists due to Assumptions \ref{jan}.
The integral representation in the top line of \eqref{theequp} follows from 
\eqref{ppre} by using the antisymmetry of the hyperbolic cotangent, the 
reality condition $\hat{G}(-\nu ) = \bigl(\hat{G}(\nu )\bigr)^{*}$ and 
$ \Im\hat{G}(\nu ) = \pi m^{-1} J(\nu )\nu |\hat{G}(\nu )|^2$.
The series representation in the bottom line of \eqref{theequp} follows from 
evaluating \eqref{ppre} with the help of the residue theorem 
and the Mittag-Leffler expansion of the hyperbolic cotangent
\begin{equation}
  \coth (\beta\hbar\nu/2) =\frac{2\nu}{\beta\hbar}\;\sum_{l=-\infty}^{+\infty}
  \frac{1}{\nu^{2}+\nu_{l}^{2}}\,.
\end{equation}
Here, $\nu _{l}:= 2\pi l/(\beta \hbar)$ are the even Matsubara frequencies. 
Note that the pole for 
$l=0$, which lies on the integration contour, counts only half due to the 
principal-value prescription.

In order for $\gesgler{q^2}$ to exist we assume $\omega > 0$ and 
start out with
\begin{equation} \label{qnmat}
\langle w^{(n)}, q^2 \rangle = \frac{\hbar}{2m} \;
\biggl(\frac{\coth(\beta \hbar\Omega /2)}{\Omega } \biggr)_{00}\,.
\end{equation}
We pick three constants $0<\varepsilon < \lambda < \tilde{\varepsilon }
< 2\pi /(\beta \hbar)$ and apply Statement \ref{omgam} to the function
\begin{equation} 
z \mapsto f(z) = \frac{z}{\lambda ^2 + z^2}\;\coth (\beta \hbar z/2)
\end{equation}
which gives
\begin{align} \label{pickq}
\langle w^{(n)}, q^2 \rangle &= \frac{\hbar}{2\pi\i m}\;
\lim_{\lambda \downto 0}\; \int_{\rz + \i\varepsilon }\!\d z \;
\frac{z^2}{\lambda ^2 + z^2}\; \hat{G}_{n}(z) \, 
\coth (\beta \hbar z /2)\notag\\
&= \frac{\hbar}{2\pi\i m}\;\lim_{\lambda \downto 0} \,\biggl\{ 
\int_{\rz + \i\tilde{\varepsilon}}\!\d z \;
\frac{z^2}{\lambda ^2 + z^2}\; \hat{G}_{n}(z)\, \coth (\beta \hbar z /2)
\notag\\ 
&\hspace*{2.3cm} + \pi \i \lambda \,\hat{G}_{n}(\i\lambda) \,
\cot (\beta \hbar \lambda /2) \biggr\} \notag\\
&= \frac{\hbar}{2\pi\i m}\;\biggl\{\int_{\rz + \i\tilde{\varepsilon}}\!
\d z \;\hat{G}_{n}(z)\, \coth (\beta \hbar z /2)\; 
+ \frac{2\pi\i}{\beta \hbar} \; \hat{G}_{n}(0) \biggr\}\,. 
\end{align}
The second equality in \eqref{pickq} follows from deforming the 
integration contour across the pole at $z=\i\lambda $. Performing the 
macroscopic limit $n\to\infty $ and subsequently the limit 
$\tilde{\varepsilon }\downto 0$, one arrives at 
\begin{equation} \label{afterpick}
\gesgler{q^2} = \frac{\hbar}{2\pi \i m}\; \valprinc\int_{\rz}\!\d\nu \;
\hat{G}(\nu )\,\coth (\beta \hbar\nu /2)\,.
\end{equation}
The integral representation in the top line of \eqref{theequq} and the 
series representation in the bottom line of \eqref{theequq} are derived 
from \eqref{afterpick} in the same way as was demonstrated for 
$\gesgler{p^2}$.
\end{proof}

\noindent
As the main result of this section, we describe the spreading of
states under the reduced averaged dynamics of Model \ref{raureidef} 
in the long-time limit.

\begin{satz} \label{klimax}
If the spectral density $ J $ of the macroscopic heat bath satisfies
Assumptions \ref{jan}, then the second moment of the particle 
momentum approaches a finite value under the reduced averaged
dynamics \eqref{reddef} of Model \ref{raureidef}
\begin{equation}   \label{rp2}
\lim_{t\to\infty } \rtad p^2 = \gesgler{p^2} + 
(-\partial_{q}^2 C)(0) \int_{\rz}\frac{\d \nu }{2\pi }\;
\nu ^2 \, \bigl| \hat{G}(\nu )\bigr|^2 \,.
\end{equation}
In the harmonically bound case $ \omega > 0 $, the same is true for the 
second moment of position 
\begin{equation}  \label{rq2}
\lim_{t\to\infty } \rtad q^2 = \gesgler{q^2} + 
\frac{(-\partial_{q}^2 C)(0)}{m^2} \int_{\rz}\frac{\d \nu }{2\pi }\;
\bigl| \hat{G}(\nu )\bigr|^2 \,.
\end{equation}
In the unbound case $ \omega =0 $  the limit \eqref{rq2} 
is to be replaced by the 
diffusive behaviour
\begin{equation} \label{rq20}
\lim_{t\to\infty } \frac{\rtad q^2}{t}  =
\frac{2}{\pi J(0)}\; \biggl( \beta ^{-1} + \frac{(-\partial_{q}^2 C)(0)}%
{2\pi J(0)}\biggr)\,.
\end{equation}
Above we have used the thermal expectation values \eqref{theequpq}.
The results imply in particular that the mean energy 
$ \rtad \bigl( p^2/(2m) + (m\omega ^2 /2) q^2\bigr) $, respectively
$ \rtad  p^2/(2m) $ in the unbound case $ \omega =0 $, 
remains finite in the long-time limit. 
\end{satz}

\begin{bems}
\item
The right-hand sides of (\ref{rp2}) -- (\ref{rq20}) are constant 
functions on phase space $\rxr\,$. Consequently, the asymptotic 
spreading of states is universal in the sense that it does not depend 
on the initial state.
\item
The terms which are due to the white noise in  
(\ref{rp2}) -- (\ref{rq20}) depend only on the curvature of its covariance 
function at the origin and, via the Green function $G$, on the
spectral density 
$ J $ of the heat bath. On the other hand, the thermal equilibrium 
expectation values $\gesgler{p^2}$ and $\gesgler{q^2}$ inhibit also a 
dependence on Planck's constant. But it is only in the subsequent
low-friction limit  
$J\downto 0$ that they reduce to the quantum-mechanical canonical
equilibrium values with respect to the system Hamiltonian alone,
see Remark \ref{zutheequ} and confer also \citet{FoKaMa65}, 
\citet{Dav73}, \citet{BeKa81} and \citet{Maa84}.
\item
In \citet{JaKu82} and \citet{Hei83a} it has already been conjectured
that the interplay 
of white noise and dissipation would lead to a diffusive spreading in
position if $\omega =0$. Eq.\ \eqref{rq20} provides the
corresponding quantitative statement.
It is to be compared with the $ t^3 $-behaviour (\ref{freiq2})
of $ \ttad q^2 $ in the situation without dissipation.
The $ \hbar $-independence of the diffusion constant on the right-hand
side of (\ref{rq20}) reveals the classical character of this phenomenon.
\item
In the absence of white noise, $C(0)=0$, Eqs.\
(\ref{rp2}) -- (\ref{rq20}) reduce to known results for the damped 
harmonic oscillator ($\omega >0$) and the damped free particle
($\omega =0$), respectively. In particluar, the
diffusion constant on the right-hand side of (\ref{rq20}) is then
given by an Einstein type of relation \citep{Sut05,Ein05,KuToHa91}. 
In the above statement we have
given precise and explicit assumptions on the spectral density under
which these results are valid. At zero temperature, $\beta = \infty$, 
or for a spectral density with $J(0)=0$ or $J(0)=\infty$, which is
excluded by our assumptions, non-exponential relaxation and anomalous 
diffusion is expected, see for 
example \citet{HaRe85}, \citet{GrScIn88} and \citet[Chap.~6]{Wei93}.
\end{bems}

\begin{proof}[Proof\phantom{x}of\phantom{x}Statement \ref{klimax}]
Statement \ref{greentotik} implies that the first term on the right-hand 
side of (\ref{redp2}) and (\ref{redq2}), respectively, does not 
contribute to the leading behaviour in the limit $ t\to\infty $.
For the case $ \omega = 0 $ the leading behaviour of (\ref{redq2})
is determined by the limit (\ref{as}). The result (\ref{rq20}) follows 
from it with the help of the relation
\begin{equation} 
\int_{\rz}\!\d t\; \varPhi_{\beta} (t) = 2\pi \beta^{-1} J(0)\,.
\end{equation}

The integral 
$ \ints \bigl( \dot{G}(s)\bigr)^2 $ in (\ref{redp2}) converges for
$ t\to\infty $. For $ \omega > 0 $  the same is true for 
the integral $ \ints \bigl( G(s)\bigr)^2 $ in (\ref{redq2}). 
Using Parseval's formula we thus conclude
\begin{equation} \label{pfou}
\int_{0}^\infty \!\d s\; \bigl( \dot{G}(s)\bigr)^2 
= \int_{\rz}\frac{\d\nu }{2\pi }\; \nu^{2}\,\bigl| \hat{G}(\nu )\bigr|^2 
\end{equation}
and, if $ \omega > 0 $,
\begin{equation} \label{qfou}
\int_{0}^\infty \!\d s\; \bigl( G(s)\bigr)^2 
= \int_{\rz}\frac{\d\nu }{2\pi }\; \bigl| \hat{G}(\nu )\bigr|^2 \,.
\end{equation}
This explains the last term in (\ref{rp2}) and (\ref{rq2}), respectively.
Finally, the convolution theorem, Definition (\ref{ankov}) and 
Statement \ref{theequ} imply
\begin{align} \label{pupsfou}
&\int_{0}^\infty \!\d s\int_{0}^\infty \!\d s' 
\dot{G}(s)\,\varPhi_{\beta} (s-s') \,\dot{G}(s')\notag\\
& \hspace*{2cm}= \frac{\hbar}{2} \int_{\rz}\!\d\nu\; J(\nu )\,
\nu^{3}\,\bigl| \hat{G}(\nu )\bigr|^2 \, 
\coth (\beta \hbar\nu /2) \notag\\
& \hspace*{2cm}= \gesgler{p^2}
\end{align}
and, if $ \omega > 0 $,
\begin{align} \label{qupsfou}
&m^{-2} \int_{0}^\infty \!\d s\int_{0}^\infty \!\d s' 
G(s)\,\varPhi_{\beta} (s-s')\, G(s')  \notag\\
& \hspace*{2cm}= \frac{\hbar}{2m^2} \int_{\rz}\!\d\nu\; J(\nu )\,
\nu\,\bigl| \hat{G}(\nu )\bigr|^2 \,
\coth (\beta \hbar\nu /2)\notag\\
& \hspace*{2cm}= \gesgler{q^2}\,.
\end{align}
\end{proof}


\section{Outlook} \label{inandout}

The present paper is devoted to the averaged dynamics of white-noise
systems. Natural extensions of this work concern the fluctuations
around the averaged dynamics and the effects of a non-vanishing
correlation time, that is, coloured-noise perturbations. Both problems
have not been treated satisfactorily so far. Diverse attempts, both
analytical (but based on uncontrolled approximations) and numerical,
can be found in the literature. To our knowledge, all of them 
treat only the case without dissipation. Unfortunately, their results are not
free from contradictions. It may therefore be of help to compile
some of these references and to relate their findings.

In the preceding sections we analyzed the averaged dynamics 
$w(p,q) \mapsto \overline{w_{t}(p,q)}$ in order to study the average
$\overline{\langle w_{t},a \rangle}$ of the quantum-mechanical
expectation value $\langle w_{t},a \rangle$ of an
observable $a$ at time $t$. Fluctuations around this average are described by
the higher moments $\overline{{\langle w_{t},a \rangle}^N}$, $N\ge 2$,
for which Jensen's inequality
\begin{equation} 
  \overline{{\langle w_{t}, a \rangle}^{N}} \ge
  \left(\overline{\langle w_{t}, a \rangle}\right)^{N}
\end{equation}
provides a (generally rather poor) lower bound. The appropriate tool
for studying the $N$-th moment $\overline{{\langle w_{t},a
    \rangle}^N}$ is the averaged dynamics of a non-interacting
$N$-particle white-noise system. In particular, it provides the
relevant information about the mapping $w(p_{1},q_{1})\cdot\ldots\cdot
w(p_{N},q_{N}) \quad\mapsto\quad
\overline{w_{t}(p_{1},q_{1})\cdot\ldots\cdot w_{t}(p_{N},q_{N})}$.
This dynamics is again given by a semigroup, whose generator can be
constructed explicitly. However, since one has not succeeded in
representing the semigroup as in Statement \ref{ttww}, which was the
starting point for all the exact calculations, one had to resort to
approximations.  \citet{Sti92} has carried out an analysis up to 
second-order perturbation theory for $N=2$ particles on a lattice with
nearest-neighbour hopping and a site-diagonal white-noise
perturbation. Taking the lattice position $\myvec{q}=(q_{1},\ldots ,q_{d})$ 
as the observable, he finds an asymptotic result which, in our
notation, would read 
\begin{equation} \label{stintzing}
  \overline{\langle w_{t}, \myvec{q}\rangle^2} \sim t^0
  \qquad\quad\text{for}\quad t\to\infty\,.
\end{equation}
This result is shown to be independent of the number $d$ of space
dimensions of the lattice. By appealing to an analogy to a classical
master equation, he even claims that \eqref{stintzing} represents the
exact asymptotic behaviour. However, \eqref{stintzing} should be
contrasted with the numerical calculations of \citet{BoToSo92} and
\citet{SaKaRe92} which suggest a result different from \eqref{stintzing}
\begin{equation} 
  \overline{\langle w_{t}, \myvec{q}\rangle^2} \sim \biggl\{
  \begin{array}{@{\;}l@{\quad}l} t^{1/2}\,, & d=1,2\\
    \ln t\,, & d=3
  \end{array} 
  \qquad\quad\text{for}\quad t\to\infty\,.
\end{equation}
Other aspects of the averaged dynamics of many-particle white-noise
systems can be found in \citet{GiMa79} and \citet{Jau87}.

Coloured-noise perturbations have attracted much attention in the last
decade. On the one hand they give rise to less idealized models, on
the other hand they provide a variety of interesting features due to
the existence of an additional time scale, the correlation time $\tau$
of the noise. Because of this complexity, there are only few exact
analytical results available. Moreover, these are limited to models in
which the noise has some additional nice properties. We only mention a
Markovian noise in time \citep{Che82,Pil85,Pil86,Tch97}, rank-one
perturbations with coloured-noise coupling coefficients
\citep{Chv91,NeSp94} and coloured-noise couplings to some
deterministic external force \citep{KiHa79,Ina81}. For classical
systems, however, there exists a more detailed picture, see for example the
review by \citet{HaJu95}. Quantum-mechanical models similar to ours
with a Gaussian coloured-noise have been studied by \citet{GoFeZe91},
\citet{Ros92}, \citet{Hei92}, \citet{LeMaFe95} and \citet{Hei96}. The
starting point for the calculations in \citet{GoFeZe91} is the
corresponding classical model which is analyzed within the
Martin-Siggia-Rose formalism. After performing a \emph{short}-time
expansion they derive the asymptotic \emph{long}-time behaviour
\begin{equation}\label{golu}
  \begin{split}
    \overline{\langle w_{t}, \myvec{p}^2\rangle} &\sim
    \biggl\{\begin{array}{@{\;}l@{\quad}l} t^{2/5}\phantom{^1}\,, & d=1\\
      t^{1/2}\,, & d\ge 2 \end{array}   \\[-.5ex]
    &  \hspace*{6cm}\text{for}\quad t\to\infty\,,  \\[-.5ex]
    \overline{\langle w_{t}, \myvec{q}^2\rangle} &\sim
    \biggl\{\begin{array}{@{\;}l@{\quad}l} t^{12/5}\,, & d=1\\
      t^{9/4}\,, & d\ge 2 \end{array} 
  \end{split}
\end{equation} 
which, according to a quasi-classical argument, should not be modified
by quantum fluctuations. \citet{GoFeZe91} also perform a numerical
simulation in order to support \eqref{golu}.  In a Comment,
\citet{Ros92} criticizes the derivation of \eqref{golu}, as it does
not guarantee the conservation of probability. His alternative
derivation, which is not presented in detail, leads to a modification
of \eqref{golu} in more than one space dimension
\begin{equation}
  \begin{split}
    \overline{\langle w_{t}, \myvec{p}^2\rangle}  &\sim
    \phantom{\biggl\{\;} t^{2/5} \quad\phantom{^{1}\,\,,  d=1}\\[-.5ex]
    &  \hspace*{6cm}\text{for}\quad t\to\infty\,.\\[-.5ex]
    \overline{\langle w_{t}, \myvec{q}^2\rangle} &\sim
    \biggl\{\begin{array}{@{\;}l@{\quad}l} t^{12/5}\,, & d=1\\
      t^{2}\,, & d\ge 2 \end{array} 
  \end{split}
\end{equation}       
A genuine quantum-mechanical treatment of the model is given by
\citet{Hei92}. He succeeds in averaging the von Neumann equation up to
first order in $\tau/t$ yielding
\begin{equation}
  \begin{split}
    \overline{\langle w_{t}, \myvec{p}^2\rangle} &\sim
    t^{\phantom{3}}
    \bigl\{ 1 + \mathcal{O}\bigl( (\tau /t)^2 \bigr)\bigr\}\\[-.5ex]
    &   \hspace*{6cm} \text{for}\quad t\to\infty\,.\\[-.5ex]
    \overline{\langle w_{t}, \myvec{q}^2\rangle} &\sim t^3 \bigl\{ 1
    + \tfrac{3}{2}\,(\tau /t) + \mathcal{O}\bigl( (\tau /t)^2
    \bigr)\bigr\}
  \end{split}
\end{equation}                            
Using a similar method, \citet{Hei96} also investigates the
corresponding behaviour at intermediate times. \citet{LeMaFe95} argue
on rather speculative grounds that \eqref{golu} should be correct up
to very long times, but in the limit $t\to\infty$ the exponents should
be replaced by smaller ones.

We conclude that the effects of a non-vanishing correlation time are
controversially discussed at present. In order to arrive at a unifying
picture, it seems that suitable bounds and other controlled approximations to
the averaged dynamics are needed.


\appendix

\section{The Weyl-Wigner-Moyal representation}

In this Appendix we intend to give a brief overview of the linear 
phase-space representation of quantum mechanics which dates back
to \citet{Wey28}, \citet{Wig32} and \citet{Moy49}.
More detailed information and proofs of the results mentioned here
can be found in the review articles \citet{Tat83}, \citet{BaJe84}
and \citet{HiOCScWi84}.
Extensions of the formalism in order to describe many-particle systems 
are dealt with in \citeauthor{GaNi93} (\citeyear{GaNi93}, 
\citeyear{GaNi94}), extensions to a 
non-Euclidean phase space in \citet{BaFlFrLi+78} 
and \citet{KaPe94}.
Some more mathematical issues are contained in \citet{Poo66}, 
\citeauthor{Dau80} (\citeyear{Dau80}, \citeyear{Dau83}),
\citet{Fol89}, \citet{Lie90} and
\citet{Fed96}. The articles \citet{MoOs94}, \citet{OsMo95},
\citet{Ara95} and \citet{Rob93}
are concerned with quasi-classical properties and the classical limit. 

For simplicity we will restrict ourselves to describe a single 
spinless quantum particle which has the Euclidean line $ \rz $ as its 
configuration space. 
The generalization to several Cartesian degrees of freedom is
merely a matter of notation.
In the standard Hilbert-space formulation
states and observables of the particle are represented by certain linear 
operators on $ \mathrm{L}^2(\rz ) $, the Hilbert space of Lebesgue 
square-integrable
complex-valued functions on $ \rz $. The link to phase space 
is provided by

\begin{definition}\label{wwabb}
Let $ F $ be a linear operator on $ \mathrm{L}^2(\rz ) $. The mapping
$ \mathcal{W} : F \mapsto \mathcal{W}(F):= f $, 
\begin{equation} \label{wwsymb}
f: \rxr \rightarrow \cz\,,\qquad\quad 
f(p,q):= \int_{\rz}\!\d r\; \e^{\i pr/\hbar}\;
\langle q-r/2 | F | q+r/2\rangle
\end{equation}
is called Weyl-Wigner mapping. The phase-space function $ f $ is called
the Weyl-Wigner symbol of the operator $ F $. 
In \eqref{wwsymb} we have used 
the Dirac notation for the integral kernel of an operator in the 
position representation.
\end{definition}

\noindent
\citet{Poo66} shows that
the Weyl-Wigner mapping induces an isometric isomorphism between the 
space of Hilbert-Schmidt operators on $ \mathrm{L}^2 (\rz) $ and the 
Hilbert space $ \mathrm{L}^2 (\rz\times\rz) $ of square-integrable 
complex-valued functions on phase space. For an extension of
(\ref{wwsymb}) to the Banach space of bounded operators on
$ \mathrm{L}^2 (\rz) $, see \citet{Dau80}. For the purpose of the present 
paper we may safely ignore these more subtle issues, and 
we will use (\ref{wwsymb}) for all operators $ F $ for which it makes 
sense, possibly in a distributional one.
 
Some important properties of the Weyl-Wigner mapping are listed without 
proof in

\begin{langsatz}
\item   \label{wweiglin}
Compatibility with the complex-linear structure:
\begin{equation}
\mathcal{W}\bigl( \alpha F +\beta G^{\adjoint}\bigr)
= \alpha \mathcal{W}(F) + \beta \mathcal{W}(G)^{*}
= \alpha f + \beta g^{*} \,,
\end{equation}
where $ \alpha ,\beta \in\cz $,  $ G^{\adjoint} $  is the Hilbert adjoint 
of $ G $ on $ \mathrm{L}^2 (\rz) $ and $ g^* $ is the complex conjugate 
of $ g $.
\item   \label{wweigprod}                                         
Operator products correspond to star products of symbols (the arrows 
indicate on which symbol a particular differentiation is meant to act):
\begin{equation}   \label{lrdiffer}
\mathcal{W}(FG) = f \exp\left\{ - \frac{\i\hbar}{2}\;\Bigl( 
\overset{\leftarrow}{\partial}_{p}\,\overset{\rightarrow}{\partial}_{q}
- \overset{\leftarrow}{\partial}_{q}\,\overset{\rightarrow}{\partial}_{p}
\Bigr)\right\}\,g =: f\star g \,.
\end{equation}
\item  \label{wweigisom}
Isometry:
\begin{equation} \label{spur}
\Sp ( F^{\adjoint} G ) =
(2\pi \hbar)^{-1} \intpr\d p\,\d q\; f^*(p,q) \; g(p,q)
=: (2\pi \hbar)^{-1} \langle f,g\rangle\,.
\end{equation}
\item 
Weyl quantization as the inverse mapping:
\begin{align} \label{weylquanten}
F = \mathcal{W}^{-1}(f) &= \intpr\,\frac{\d p\d q}{2\pi \hbar}\;
\e^{\i (p- P)q /\hbar}\; f(p, Q +q/2) 
\notag\\
&= f(-\i\hbar\partial_{q'}, -\i\hbar\partial_{p'})\, 
\left. \e^{\i (q'P + p'Q)/\hbar} \right|_{p'=0=q'}\,.
\end{align}
Here $ P $ and $ Q $ denote the momentum and position operator, 
respectively.
\item    \label{wweigranddicht}
Diagonal matrix elements by integration:
\begin{equation}
\int_{\rz}\frac{\d p}{2\pi \hbar}\; f(p,q)  = 
\langle q | F | q\rangle\,,\qquad
\int_{\rz}\frac{\d q}{2\pi \hbar}\; f(p,q)  = 
\langle p | F | p\rangle\,.
\end{equation}
\end{langsatz} 

\begin{bem} \label{moyalbem}
According to Statement \ref{wweigprod} 
the symbol of the standardized commutator is given by 
\begin{align}\label{moyal}
\mathcal{W}\bigl((\i/\hbar) (FG - GF)\bigr)
&= (\i/\hbar) (f\star g -g\star f) \notag\\
&= f\; \frac{2}{\hbar}\,\sin\left\{ \frac{\hbar}{2}\;\Bigl( 
\overset{\leftarrow}{\partial}_{p}\,\overset{\rightarrow}{\partial}_{q}
- \overset{\leftarrow}{\partial}_{q}\,\overset{\rightarrow}{\partial}_{p}
\Bigr)\right\}\,g \notag\\
&=: [f,g]\,.
\end{align}
It is called Moyal bracket. If either $ f $ or $ g $ is a quadratic 
polynomial in $ p $ and $ q $, then the Moyal bracket reduces to the 
Poisson bracket
\begin{equation} \label{poisson}
[f,g] = (\partial_{p}f)(\partial_{q}g) - (\partial_{q}f)(\partial_{p}g)\,.
\end{equation}
Regardless of this assumption on $ f $ and $ g $, the equality 
(\ref{poisson}) is always true in the classical limit $ \hbar\downto 0 $.
\end{bem}

There are uncountably many ways of associating a phase-space function 
with a quantum-mechanical operator and vice versa
\citep{Coh66,AgWo70,FiLe93}. The Weyl-Wigner mapping, respectively the 
Weyl quantization, provides the one that treats position and momentum
in a totally symmetric way. This is illustrated by the following

\begin{bsple}
\item
$ \disp \mathcal{W}^{-1}\bigl(f(q)\bigr) =f(Q)\,,
\quad \mathcal{W}^{-1}\bigl(f(p)\bigr) =f(P) $,
\item
$ \disp\mathcal{W}^{-1}\bigl( pf(q)\bigr) = \tfrac{1}{2}\, 
\bigl(P f(Q) + f(Q)P\bigr) $,
\item
$ \disp \mathcal{W}\bigl(P f(Q)\bigr) = pf(q) -
(\i\hbar/2)\,(\partial_{q}f)(q) $,
\item
$ \disp \mathcal{W}\bigl(Pf(Q)P\bigr) = p^2 f(q) +
(\hbar^2/4)\,(\partial_{q}^2 f)(q) $.
\end{bsple}

\noindent
Phase-space functions which represent quantum-mechanical states are 
called Wigner densities. By convention we say that a 
phase-space function $ w $ is a Wigner density if it can be written as
\begin{equation}\label{widichte}
w = (2\pi \hbar)^{-1}\,\mathcal{W}(W)\,,
\end{equation}
where $ W $ is some state operator, that is, $ W = W^{\adjoint} \ge 0 $ and
$ \Sp W =1 $.

\begin{satz} \label{wigprop}
Wigner densities are bounded, $ |w(p,q)| \le (\pi \hbar)^{-1} $,
normalized, $ \langle w,1\rangle =1 $, and 
square-integrable, $ \langle w,w\rangle \le (2\pi \hbar)^{-1} $. 
Moreover, one has $ \langle w,w\rangle  =  (2\pi \hbar)^{-1} $ 
if and only if $ w $ represents a pure state.
\end{satz}

\noindent
In contrast to probability densities, however, Wigner densities 
may also take on negative values.
This fact is a consequence of Statements \ref{wweiglin} and 
\itemref{wweigranddicht}, see e.g.\ \citet{Wig71}. A simple example is 
provided by the pure state $ W= |\psi  \rangle\langle \psi | $ 
whose position amplitude $ \langle q|\psi \rangle $ is the positive 
square root of a uniform probability density over some 
interval. This example shows in addition that Wigner densities need not be 
Lebesgue integrable \citep{Dau83}.

Below we give two intrinsic characterizations of 
Wigner densities. The first one relies on the factorization
$ W=\bigl(W^{1/2}\bigr)^2 $  \citep{Shi79}, the second 
one reflects the corresponding properties of state operators 
\citep{NaOC86}.

\begin{satz}\label{wdchar0}
A phase-space function $ w $ is a Wigner density, if and only if
there exists a real-valued phase-space function $ u $ 
with $ \langle u ,u \rangle =1 $ such that $ w=u \star u $.
\end{satz}

\begin{satz}\label{wdchar}
A phase-space function $ w $ is a Wigner density, if and only if it has
the following three properties:
\begin{enumerate}
\item[\textup{1.}] $ w=w^* $,
\item[\textup{2.}] $ \langle w,1\rangle=1 $,
\item[\textup{3.}] $ \langle w,v\rangle \ge 0 $ ~~for all Wigner densities 
$ v $.
\end{enumerate}
\end{satz} 

\noindent
Statement \ref{wdchar0} supports that Wigner densities 
converge to probability densities in the classical limit 
$ \hbar \downto 0 $ (at least in some weak sense). From a rigorous 
mathematical point of view, however, such questions seem to be highly 
non-trivial. This is due to severe technical difficulties which arise in 
part from the missing Lebesgue integrability of Wigner densities.
Related aspects of the classical limit of quantum-mechanical phase-space 
representations may be found in \citet{Rob93}, \citet{Ara95}
and \citet{Wer95}.


\begin{acknowledgements}
  It is a pleasure for Hajo Leschke to thank Wolfgang Kundt for
  guiding his first steps in theoretical physics with so much
  enthusiasm. In particular, he explained to him the Weyl-Wigner-Moyal
  representation some thirty years ago \citep{Kun67}. This work was
  supported by the Deutsche Forschungsgemeinschaft under grant number
  Le 330/10--1.
\end{acknowledgements}



\end{document}